\documentclass[12pt]{amsart}
\usepackage{amssymb, amsthm, amsfonts, latexsym}
\usepackage[leqno]{amsmath} 
\usepackage{natbib}
\usepackage{mathtools}
\usepackage{enumitem}
\usepackage{needspace}
\usepackage{array}
\usepackage{graphicx,booktabs,tabularx,longtable,pdflscape,float}
\usepackage{xurl}
\usepackage[multiple]{footmisc}
\usepackage{comment}

\usepackage[left=1.25in, top=1.25in, right=1.25in, bottom=1.25in]{geometry}
\usepackage{setspace}
\usepackage{tgpagella}
\usepackage{microtype}

\usepackage{xcolor}
\usepackage{tikz}

\usepackage{etoolbox}
\patchcmd{\section}{\scshape}{\bfseries}{}{}
\makeatletter
\renewcommand{\@secnumfont}{\bfseries}
\makeatother

\theoremstyle{plain}
\newtheorem{theorem}{Theorem}

\newtheorem{lemma}{Lemma}

\newtheorem{proposition}{Proposition}
\newtheorem*{proposition*}{Proposition}
\newtheorem{claim}{Claim}
\newtheorem{example}{Example}

\theoremstyle{definition}
\newtheorem{definition}{Definition}

\newtheoremstyle{boldremark}
 {0.5\topsep}{0.5\topsep}
 {\normalfont}{}
 {\bfseries}{.}
 {.5em}{}
\theoremstyle{boldremark}
\newtheorem{remark}{Remark}

\usepackage[
 colorlinks=true,
 linkcolor=blue,
 citecolor=blue,
 urlcolor=blue,
 filecolor=blue,
 hyperfootnotes=false,
 hyperindex,
 breaklinks
]{hyperref}

\usepackage{url}
\begin{document}
\raggedbottom
\author[Aygün]{Orhan Aygün$^\textrm{1}$}\thanks{$^\textrm{1}$ Department of Economics, Boğaziçi University. Email: \href{orhan.aygun@bogazici.edu.tr}{orhan.aygun@bogazici.edu.tr}}

\author[Turhan]{Bertan Turhan$^\textrm{2}$}\thanks{$^\textrm{2}$ Department of Economics, Iowa State University. Email: \href{bertan@iastate.edu}{bertan@iastate.edu}}

\date{\today}

\title{Reconciling Affirmative Action and Merit in India: An Axiomatic Analysis}\thanks{\textit{Use of AI}: The first version of this paper was circulated in 2021. It has been revised multiple times since then. We used ChatGPT and Claude to search related court rulings in the Indian judiciary, polish the paper's writing, and verify the proofs. All ideas and scientific content were developed by the authors. We take full responsibility for the content in this paper. \\
This paper replaces and supersedes \cite{aygun2024over} and \cite{aygun2024tale}.
We especially thank Ibrahim Suat Evren, Isa Hafalir, Manshu Khanna, Onur Kesten, and Bumin Yenmez for their suggestions. We also thank Inácio Bó, Battal Dogan, Umut Dur, Kenzo Imamura, and Kriti Manocha, as well as the seminar and conference participants at Bogazici University, Washington University in St Louis, ISI Delhi, University of Iowa, the Greater Bay Area Market Design Workshop 2024 and 2026, the Behavioral and Experimental Economics Conference at ITAM (in memory of Piotr Evdokimov), the 4th Durham Economic Theory Conference, the 17th Matching in Practice Workshop, the Kansas CARE-Colloquium, the Kansas Workshop in Economic Theory 2024, and the Stony Brook Game Theory Festival 2026 for their helpful feedback.}

\begin{abstract}
\onehalfspacing

How should affirmative action policies reconcile legally mandated reservations with merit-based allocation? We address this question through an axiomatic analysis of India's reservation system, tracing its legal evolution through three landmark Supreme Court decisions. We formalize the rules governing vertical reservations, eligibility cutoffs, and the transfer of unfilled reserved positions to the open category and characterize the resulting assignment rules. We show that endogenous eligibility cutoffs can generate inefficiencies, incompatibilities among legal requirements, and violations of substitutability, whereas the exogenous cutoff interpretation established in P.V. Indiresan (2011) admits a family of implementable rules. Our main result establishes that forward transfer of vacant reserved positions with exogenous cutoffs merit-dominates all other rules in this family. Moreover, every rule in the family satisfies substitutability and size monotonicity, making it compatible with deferred acceptance. Our findings highlight cutoff rules and transfer sequencing as key policy instruments for reconciling legal requirements with merit-based selection.

\end{abstract}
\maketitle

\newpage
\section{Introduction}

In many high-stakes environments, such as school choice, college admissions, and public-sector recruitment, allocative authorities must reward performance while also expanding access for historically excluded groups. These objectives are often treated as competing ethical ideals, but from a design perspective, they are also competing operational constraints. Once diversity goals are encoded through quotas, reserves, or priority rules, they interact in nontrivial ways with merit-based rankings and how available positions are filled. The relevant economic problem is therefore not whether diversity should come at the expense of meritocracy in the abstract, but \textit{how institutional rules can reconcile these goals without generating avoidable inefficiency, arbitrariness, or legal incoherence}.\footnote{There is now a large literature in market design studying diversity and affirmative action for real-world object allocation problems. Examples include \cite{hafalir2013effective} for minority reserves and majority quotas, \cite{echenique2015control} for diversity-sensitive choice rules, \cite{kamada2015efficient} for Japanese residency matching, \cite{kominers2016matching} for slot-specific priorities and reserves, \cite{dur2018reserve} for the policy impact of reserve precedence in Boston's school choice program, \cite{DelacretazKominersTeytelboym2023} for matching mechanisms for refugee resettlement and \cite{PRS2025} for immigration lottery design among many others.} 

This tension is especially sharp in India, where the legal framework simultaneously seeks to remedy historical disadvantage and preserve standards of excellence. In this paper we take the binding legal requirements as the primitive of the design exercise and formalize the mandates and directives arising from three landmark stages of Supreme Court jurisprudence---\textit{Indra Sawhney v. Union of India (1992), Ashoka Kumar Thakur v. Union of India (2008) together with a clarification order on the ruling, and P.V. Indiresan v. Union of India (2011)}---as mathematical restrictions on assignment rules and ask which allocation procedures implement them.\footnote{These judgments can be found at \href{https://indiankanoon.org/doc/1363234/}{Indra Sawhney (1992)}, \href{https://indiankanoon.org/doc/63489929/}{Ashoka Kumar Thakur (2008)}, and \href{https://indiankanoon.org/doc/1340286/}{P.V. Indiresan (2011)} (last accessed on 11/14/2026). Paragraph 5 of the \href{https://indiankanoon.org/doc/1340286/}{P.V. Indiresan (2011)} judgment reproduces the 14 October 2008 clarificatory order on the Ashoka Kumar Thakur (2008) ruling, while the subsequent paragraphs explain its interpretation.}

\subsection{Institutional and Legal Background}
Reservation policies in India were initially implemented to ensure the representation of historically discriminated Scheduled Castes (SC) and Scheduled Tribes (ST) in public education and employment. Articles 15(4) and 16(4) of the Indian Constitution provided reservations for SCs and STs in educational institutions and government jobs. In 1954, the Ministry of Education suggested a 20\% reservation for SCs and STs in educational institutions, with 15\% for SCs and 5\% for ST. In 1982, the percentage of SC and ST reservations increased to 15\% and 7.5\%, respectively, in central government jobs.

Reservations for Other Backward Classes (OBCs) were introduced after the recommendations of the Mandal Commission, which proposed a 27\% reservation for OBCs in government employment. The resulting policy was challenged before the Supreme Court of India (SCI) multiple times, leading to landmark rulings. The relevant legal framework was subsequently developed in \textbf{three stages}.

Initially, the 27\% reservation for Other Backward Classes (OBCs) was challenged before the SCI, culminating in its landmark judgment in \textit{Indra Sawhney v. Union of India} (1992), which upheld the 27\% OBC reservation and established the foundational framework governing vertical reservations.\footnote{\cite{sonmez2022affirmative} analyze the Indra Sawhney (1992) judgment from a matching-theoretic perspective.} The unreserved positions are \textbf{open-category} positions and are accessible to all applicants, including applicants who belong to a vertical reserve category. Applicants who do not identify with SC, ST, or OBC are conventionally described as belonging to the \textbf{general category}. The judgment introduced three fundamental principles.

First, the \textbf{over-and-above} (OA) principle requires the highest-scoring applicants to be admitted through the open-category, irrespective of their reserve-category membership. Second, \textbf{within-category fairness} (WCF) requires respect for merit within each category. Third, \textbf{quota-filling subject to eligibility} (QFE) requires each reserve to be filled whenever eligible applicants remain. The three principles  characterize the following simple sequential process to implement vertical reservations introduced in Indra Sawhney (1992): 
\begin{quotation}
    \textit{fill the original open-category positions by overall merit and then fill each reserve category by within-category merit.} 
\end{quotation}

In the second stage, the 27\% OBC reservation under the \textbf{2006 Central Educational Institutions Act}, also known as the \textbf{Reservations in Admissions Act}, was challenged. In \textit{Ashoka Kumar Thakur (2008)}, the Constitution Bench upheld the 27\% OBC reservation in central educational institutions. The judgment emphasized that reservation should be reconciled with \textbf{standards of excellence} as follows.

\begin{quote}
  \textit{To maintain standards of excellence, cut off marks for OBCs should be set not more than 10 marks out of 100 below that of the general category}.\footnote{There is widespread confusion in India about the types of positions and the types of applicants. A cursory search on the Internet reveals that the term ``general seats" or "general category" are used to describe open-category positions. The same confusion also arises in SCI verdicts. In the above quote, by ``general category," they mean the ``open-category." \cite{aygun2022dereserve} shows that there are significant unintended consequences if the remaining OBC positions are \textbf{only} provided to students in the general category. Consistent with our interpretation, vacant OBC positions have been assigned as an open-category in admissions to technical universities. See \cite{baswanaetal2019} for further details.}   
\end{quote}

A subsequent \textbf{clarificatory order dated October 14, 2008}, stated that 
\begin{quote}
    \textit{the OBC cutoff should be 10\% below the cutoff for open-category candidates.} 
\end{quote}

The clarificatory order \textit{instructed} central educational institutions to fill vacant OBC seats with other eligible students while observing inter se merit.\footnote{The chronology and the text of the October 14, 2008 clarification are reproduced and interpreted by the Supreme Court in \textit{P.V. Indiresan v. Union of India} (2011).} The issue of OBC vacancies was addressed in the clarificatory order as follows.

\begin{quote}
    \textit{We are told that in many of the Central Educational Institutions the seats which are to be filled up by OBC candidates are still remaining vacant. These institutions may endeavor to fill up these vacant seats by other eligible students at the earliest, i.e at least by the end of October 2008 observing inter se merit of the candidates.} 
\end{quote}

The phrase ``\textbf{cutoff marks}'' in Ashoka Kumar Thakur (2008) and in the clarificatory order was ambiguous. It was interpreted in two different ways: \textit{endogenous interpretation} and \textit{exogenous interpretation}.  

 Under the \textbf{endogenous cutoff} interpretation, the relevant benchmark is endogenous in the sense that the OBC threshold is determined from the score of the last applicant admitted to an open-category position. When de-reservation occurs due to vacancies in the OBC category, the definition of the "last applicant admitted to an open-category" becomes ambiguous. There are two potential interpretations in the endogenous cutoff case, depending on whether reverted vacant OBC positions impact the open-category cutoff or not. 

\begin{enumerate}
    \item \textbf{Original-open cutoff interpretation:}
    The cutoff is determined by the score of the last
    applicant admitted to an original open-category position,
    excluding positions created through de-reservation.

    \item \textbf{Expanded-open cutoff interpretation:}
    The cutoff is determined by the score of the last
    applicant admitted to an open-category position,
    including positions created through de-reservation.
\end{enumerate}

In 2008-2009 and 2009-2010, \textbf{Jawaharlal Nehru University (JNU)} implemented a procedure where the OBC threshold was determined only after all open-category seats were filled. In particular, although JNU reverted vacant OBC seats to open-category and allocated them in order of merit, it applied the first interpretation—defining the threshold using the last candidate admitted to an original open-category position. The JNU interpretation makes OBC eligibility depend on the realized applicant pool and on the allocation of open-category positions. In P.V. Indiresan, the Court explained the JNU practice as follows.

\begin{quote}
    \textit{The Jawaharlal Nehru University (for short `JNU'), second respondent herein, interpreted the said order of this Court dated 14.10.2008 to mean that the minimum marks for admission to be secured by an OBC candidate should not be less than the marks secured by the last student admitted under general category less 10\%. The admissions for 2008-09 and 2009-10 were done on that basis. As a result, it would appear considerable number of OBC seats got reverted to general category for non-availability of eligible OBC students with the required marks.}
\end{quote}

The JNU implemented an endogenous cutoff rule under which the OBC cutoff was determined by the score of the last applicant admitted to an original open-category position, while permitting the de-reservation of unfilled OBC positions. 

In contrast, in designing the admission mechanism for engineering colleges in India, \cite{baswanaetal2019} adopted the alternative interpretation, under which the open-category cutoff is determined by the score of the last applicant admitted to an open-category position, including positions transferred from the OBC category through de-reservation.\footnote{Indeed, this interpretation provides the rationale for the design team's use of \textbf{backward transfers}: unfilled OBC positions are transferred to the open category, which precedes the OBC category in the allocation sequence, and the deferred acceptance algorithm is rerun with the updated capacities. This process is repeated until no further OBC vacancies remain. The stated policy objective for this decision was to guarantee a higher cutoff score for open category than those of reserved categories.}

Under the \textbf {exogenous cutoff} interpretation, the relevant benchmark is \textbf{exogenous} in the sense that ``cutoff'' means the minimum eligibility or qualifying score \textbf{announced in advance} for open-category positions, and the OBC minimum is permitted to be lower only by the amount allowed by the Court. The exogenous interpretation fixes the eligibility of the OBC before the assignment begins.

The third stage resolves the cutoff score ambiguity. In \textit{P.V. Indiresan v. Union of India} (2011), the SCI held that the expression ``cutoff marks'' in the October 14, 2008 order referred to the \textit{minimum eligibility/qualifying marks} prescribed in advance, not to the score of the last candidate admitted in the open-category. The Court rejected JNU's procedure as ``arbitrary and discriminatory'' because it made eligibility depend on an uncertain benchmark determined only after applications had been submitted.\footnote{See \textit{P.V. Indiresan v. Union of India}, (2011) 8 SCC 441, paras. 32--33.}

The Court also made the reversion of vacant OBC seats a requirement in P. V. Indiresan (2011). The SCI stated in paragraph 41(ii) that 

\begin{quotation}
    \textit{Only if OBC candidates possessing the minimum eligibility/qualifying marks are not available in the OBC merit list, the OBC seats shall be converted into general category seats.}
\end{quotation}

An important recent confirmation on OBC de-reservation appears in the Delhi High Court's judgment of 18 May 2026, National Board of Examinations v. Dr. Aditi Panwar and Others.\footnote{The case is available at https://indiankanoon.org/doc/144626216/ (last accessed on 11/18/2026).} In paragraph 42, the High Court expressly discusses P. V. Indiresan and explains that OBC-reserved seats remaining vacant after the applicable qualifying requirements have been applied may be filled by unreserved-category candidates. Paragraph 43 also discusses the principle that reserved seats should not be wasted once eligible reserved-category applicants have been exhausted.

\subsection{Legal Status of Cutoffs}
Justice Bhandari first addresses setting reasonable cut-off marks and reverting unfilled seats to the general category in paragraph 371 of his opinion in Ashoka Kumar Thakur (2008), noting that 
\begin{quote}
     \textit{Reasonable cut off marks should be set so that standards of excellence are not greatly affected. The unfilled seats should revert to the general category.}
\end{quote}

The "10\%" language later entered litigation practice through the October 14, 2008 clarificatory order in P.V. Indiresan stating that the maximum cut-off marks for OBCs be 10\% below general category cut-off marks. It was subsequently interpreted in P.V. Indiresan (2011) as referring to relaxation on minimum eligibility/qualifying marks rather than a rigid gap.

\subsection{Legal Status of OBC De-reservation}
In Ashoka Kumar Thakur (2008), Justice Bhandari expressed the view that unfilled OBC seats should revert to the open category. In his individual opinion, this was a judicial observation/recommendation rather than a mandatory direction issued by the full Constitution Bench.

In the 14 October 2008 Clarificatory Order, the SCI addressed vacant OBC seats in Central Educational Institutions, stating that institutions may endeavor to fill up these vacant seats by other eligible students at the earliest, i.e., at least by the end of October 2008 observing inter se merit of the candidates.

The legal status of OBC de-reservation became a binding rule in P.V. Indiresan (2011). The SCI clarified and made reversion operative as a conditional rule. This principle has also been reaffirmed in later High Court decisions, such as the Delhi High Court's May 18, 2026 judgment in National Board of Examinations in Medical Sciences v. Dr. Aditi Panwar \& Ors.)

\section*{Overview of Results}

Our primary contributions are threefold. First, we analyze the historical evolution of OBC reservations, focusing on the rules governing OBC cutoffs and the reversion of unfilled OBC seats. Second, we translate the Court's mandates and directives into explicit mathematical axioms, providing formal characterizations of the allocation rules. Finally, reflecting jurisprudence's strong emphasis on maintaining the standard of excellence, we compare allocation rules that satisfy the legal axioms based on merit, using a version of Gale domination (\cite{gale1968optimal}), and offer concrete policy recommendations.

We begin by formalizing the principles established by Indra Sawhney (1992). The OA principle, WCF, and QFE jointly characterize the Indra Sawhney assignment rule, $\eta^{IS}$ (Theorem \ref{thm:IS}), which serves as the hard-reserve benchmark for the subsequent legal developments.\footnote{\cite{sonmez2022affirmative} also provide a similar characterization in plain English without formalization and proof. Notably, they provided a verbal description of the OA principle that differs from our definition.} While the OA principle and WCF have remained foundational requirements throughout these developments, later rulings have modified QFE by altering the eligibility criteria for OBC positions and permitting the transfer of unfilled OBC positions to the open-category.

Next, we formalize the ambiguity arising from Ashoka Kumar Thakur (2008) and its subsequent clarification of October 14, 2008. We first consider the endogenous cutoff interpretation in the absence of de-reservation, which we capture through the modified quota-filling condition (mQFE). Theorem \ref{thm:AKT} establishes that the OA principle, WCF, and mQFE jointly characterize the assignment rule $\eta^{AKT}$, under which OBC positions are treated as hard reserves. 

The assignment rule $\eta^{AKT}$ can leave many OBC positions vacant. The clarification of October 14, 2008, however, introduced an additional requirement: unfilled OBC positions must revert to the open-category and be made available to other eligible applicants. This de-reservation mandate marks a departure from the hard-reserve framework and necessitates a further modification of the assignment rule.

The introduction of de-reservation under endogenous eligibility gives rise to a fundamental design problem. If the reverted OBC positions enter the computation
of the open-category cutoff, de-reservation can alter the eligibility of the very applicants for whom these positions were originally reserved. We establish that
this feedback makes the joint implementation of the cutoff mandate, the de-reservation requirement, and the foundational allocation principles impossible
(Theorem~\ref{thm:impossibility}). This impossibility motivates the analysis of the alternative endogenous cutoff interpretation adopted by JNU.

The JNU rule, denoted by $\eta^{JNU}$, implements de-reservation by reverting unfilled OBC positions to the open-category and allocating them in a final stage. Thus, $\eta^{JNU}$ extends $\eta^{AKT}$ by introducing an additional stage in which the remaining OBC vacancies are filled based on merit scores. We refer to this de-reservation procedure as \textbf{forward transfers}. We establish that placing the reverted positions at the end of the precedence order yields the merit-optimal implementation of de-reservation (Theorem \ref{thm:etaJNU}). Despite its merit-optimality, however, the JNU rule violates the desirable substitutes property, although it satisfies consistency and size monotonicity (Proposition \ref{prop:etaJNU}).

Finally, we turn to the legally operative framework established by \textit{P.V. Indiresan (2011)}. The Court requires that the predetermined minimum eligibility score for OBC positions comply with the permissible relaxation relative to the minimum eligibility score for open-category positions. Consequently, the set of eligible OBC candidates is determined exogenously prior to the allocation process. Any OBC positions that remain vacant after the allocation to eligible OBC candidates are then reverted to the open-category. With eligibility fixed ex ante, the remaining design discretion concerns the placement of these converted open-category positions in the precedence order. In particular, the allocation procedure must determine at which stage the reverted positions are processed relative to the original open-category and reserved-category positions.

We introduce a family of assignment rules, $\mathbb{A}$, that consists of sequential assignment rules that implement OBC de-reservation under exogenously
determined eligibility thresholds. Its members share the same initial reserve capacities and merit-based selection criteria but differ in the placement of de-reserved OBC positions within the precedence order. Each assignment
rule is associated with a predetermined transfer scheme that specifies how vacant OBC positions are redistributed across different stages of the allocation sequence. Whenever OBC vacancies arise, the capacities are updated according to the transfer scheme and the allocation procedure is repeated until no genuine OBC position remains vacant. Transfer schemes are cumulative and irreversible in the sense that positions previously transferred to a particular stage cannot be subsequently relocated to another stage. Within this family, we identify the forward-transfer (FT) rule, $\eta^{FT}$, and show that it merit-dominates every other admissible transfer scheme (Theorem \ref{thm:FT}). Moreover, we establish that the choice rules induced by the members of $\mathbb{A}$ satisfy consistency, substitutability, and size monotonicity, ensuring their compatibility with the deferred acceptance (DA) algorithm (Theorem \ref{thm:family}). Establishing these results requires analytical arguments that accommodate simultaneous changes in applicant populations and endogenously determined capacity vectors, rather than comparisons of sequential allocation procedures under fixed capacities.

The two legal interpretations also admit a direct comparison. The JNU rule, $\eta^{JNU}$, employs an endogenous eligibility threshold for OBC, while the forward-transfer rule, $\eta^{FT}$, applies the exogenous threshold mandated by P.V. Indiresan (2011). Whenever the JNU threshold is weakly more stringent than the P.V. Indiresan (2011) threshold for a given applicant pool, $\eta^{JNU}$ transfers weakly more OBC positions to merit-based open competition. If the resulting allocations differ in their selected cohorts, $\eta^{JNU}$ merit-dominates $\eta^{FT}$. 

The remainder of the paper is organized as follows. Section \ref{sec:model} introduces the model and the relevant properties of the assignment and choice rules. Section \ref{sec:IS} formalizes the principles established in Indra Sawhney (1992) and provides an axiomatic characterization of the resulting assignment rule. Section \ref{sec:AKT} examines the endogenous cutoff interpretation of Ashoka Kumar Thakur (2008) in the absence of de-reservation. Section \ref{sec:endcutoff} analyzes two alternative implementations of endogenous cutoffs with de-reservation, establishing an impossibility result under one interpretation and identifying a merit-optimal rule that violates substitutability under the other. Section \ref{sec:adaptive} studies the exogenous eligibility interpretation established in \textit{P.V. Indiresan} (2011), introduces a family of iterative capacity-transfer assignment rules, and establishes the merit-optimality of forward transfers and the DA compatibility of the induced choice rules. Section \ref{sec:comparelegal} compares the two eligibility interpretations. Section \ref{sec:rellit} discusses the related literature and Section \ref{sec:concl} concludes. Proofs are provided in the Online Appendix.

\section{Preliminaries}\label{sec:model} 

We consider an institution with $\overline{q}$ positions. Let $R=\left\{ SC,ST,OBC,EWS\right\}$ be the set of reserve categories. We denote by $q^{r}$ the number of positions reserved for category $r\in R$ such that $\sum_{r\in R}q^{r} <\overline{q}$. We let $q^{o}=\overline{q}- \sum_{r\in R}q^{r}$ be the number of open (unreserved) positions. The vector $\mathbf{q}=\left(q^{o},q^{SC},q^{ST},q^{OBC},q^{EWS}\right)$ denotes the initial allocation of positions in categories. Let $I$ denote the set of all applicants.

The function $t:I\rightarrow \{g\}\bigcup R$ denotes the category membership of the applicants, where $g$ denotes the \textbf{general category}. $t(i)=g$ means that individual $i$ does not belong to any vertical reserve category, or she did not report her category membership. In India, it is $\mathbf{optional}$ to report membership in the reserve categories. Members of the reserve categories who do not report their membership are considered general category applicants and are eligible only for open-category positions. Members of the reserve category $r\in R$ are eligible for both the open-category and reserved category $r$ positions. We let $T=\left(t(i)\right)_{i\in I}$ denote the profile of the applicants' reserve category membership. Let $\mathcal{T}$ denote the set of all profiles of the applicants' reserve-category membership.

The function $\theta: I \rightarrow \mathbb{R}_{+}$ denotes the test scores of the applicants. $\theta (i) > \theta (j)$ means that applicant $i$ has a higher test score than that of $j$. Applicants' merit scores induce a priority ranking for the institution. We denote this linear order over $I\cup\{\emptyset\}$ by $\succ$.\footnote{In India, when two or more applicants have the same score, ties are broken with a given objective criterion.} Throughout the analysis, we assume that merit scores
are distinct. Thus, for any two distinct applicants
$i,j\in I$, we have
\[
i\succ j
\quad\Longleftrightarrow\quad
\theta(i)>\theta(j).
\]
 
$i\succ\emptyset$ means that the applicant $i$ is acceptable for the institution. Similarly, we write $\emptyset\succ i$ to say that applicant $i$ is unacceptable for the institution. 

For each $r\in R$, the category-specific priority ranking $\succ_r$ is defined as follows:
\begin{enumerate}
    \item For every $i\in I$ such that $t(i)=r$,
    we have $i\succ_r\emptyset$.
    \item For every $i\in I$ such that $t(i)\neq r$,
    we have $\emptyset\succ_r i$.
    \item For every $i,j\in I$ such that
    $t(i)=t(j)=r$, we have
    $i\succ_r j$ if and only if $i\succ j$.
\end{enumerate}
The ranking among applicants who are ineligible
for category-$r$ positions is immaterial.

Given the set of applicants $A$ and applicant $i\in A$, let $rank_{A}(i)$ be the rank of applicant $i$ in the set $A$ with respect to the merit ranking $\succ$. That is, $rank_{A}(i)=k$ if and only if
$\mid\left\{ j\in A\mid j\succ i\right\} \mid=k-1$.
Let $A^{r}=\left\{ i\in A\mid i\succ^{r}\emptyset\right\} $ be the set of applicants in $A$ who are eligible for category $r\in R$.

\subsection{Institutions' Choice Rules}

Conventionally, institutions' selection from a given set of applicants under affirmative action constraints is modeled via \textit{choice rules}. Choice rules, which select a subset from a given set of applicants, are insufficient for the problems we consider because they do not specify the position types under which applicants are selected. 

 Mathematically, a choice rule is a function $C: 2^{I} \longrightarrow 2^{I}$ such that, for all $A\subseteq I$, $C(A)\subseteq A$ and $\mid C(A)\mid \leq \overline{q}$, where $\overline{q}$ denotes the number of positions. A choice rule $C$ satisfies the \textbf{substitutes property} if,  for all $i,j\in I$ and $A\subseteq I \setminus \{i,j\}$,
	\[
	j\notin C(A\cup\{j\})\;\Longrightarrow\;j\notin C(A\cup\{i,j\}).
	\]

A choice rule $C$ is \textbf{consistent} if, for any $A\subseteq I$ and any $i\in I \setminus A$,  
$i\notin C(A\cup \{i\}) \text{ implies } C(A)=C(A\cup \{i\})$.

A choice rule $C$ satisfies \textbf{size monotonicity} if, for all sets of individuals $A\subseteq I$ and all individuals $i\in I \setminus A$, we have 
	$\mid C(A)\mid\leq\mid C(A \cup \{i\}) \mid$.\footnote{It is well-known that size monotonicity guarantees the strategy-proofness of the deferred acceptance mechanism when institutions' choice rules satisfy it along with substitutability.}

\subsection{Institutions' Assignment Rules}

The choice rules tell only the set of admitted applicants from a given set. However, they do not specify the position types under which they are selected. In India, the type of positions to which applicants are assigned is made public. As such, we define an \textit{assignment rule} that specifies the type of position to which individuals are assigned.  

\begin{definition}\label{def:assignmentrule}
An \textbf{assignment rule}  $\eta$ is a function where, given a set of applicants $A$ and a category membership profile $T$,  
 $$\eta(A,T) = \{(i,\tau)| i\in A  \text{ and } \tau \in \{o\}\cup R \} \text{ such that }$$
 
	\begin{enumerate}
        \item [(i)] for any $i\in A$, we have at most one $(i,\tau)\in \eta(A,T)$, where
		\[
		\begin{cases}
			\begin{array}{c}
				\tau=o \\
				\tau \in\left\{ t(i),o\right\}  
			\end{array} & \begin{array}{c}
				\text{if} \;t(i)=g,\\
				\text{if} \;t(i)\in R,
		\end{array}\end{cases}
		\]
		\item [(ii)] $\lvert \eta(A,T) \lvert \leq \overline{q}$, and
        \item [(iii)] $\mid\left\{ (i,\tau)\in \eta(A,T): \tau=r \right\} \mid\leq q^{r}$, for each $r\in R$.		
	\end{enumerate}
\end{definition}

Condition (i) states that each individual is assigned at most one position: either an open-category position or, if she belongs to a reserve category, a position in the reserve category for which she qualifies. Condition (ii) states that the total number of individuals assigned cannot exceed the total number of positions available. Condition (iii) states that the number of individuals assigned to each reserve category cannot exceed the number of positions reserved for that category. 

Each assignment rule $\eta$ induces an associated choice rule $C^{\eta}$, which specifies a selected subset of any given set of applicants. That is, for any given $A\subseteq I$, $C^{\eta}\left(A\right)\subseteq A$. Formally, given a set of applicants $A\subseteq I$, a category membership profile $T$, and an assignment rule $\eta$, the \textbf{induced choice rule} $C^{\eta}$ is defined as
\[
C^{\eta}(A)=\{i\in A \lvert \exists (i,\tau)\in \eta(A,T) \text{ for some } \tau\in R\cup \{o\}\}.
\]

\subsection{Merit Comparison}

Following the tradition of \cite{gale1968optimal}, we use the following criterion to compare two sets of individuals based on merit. 

\begin{quotation}
    For $X, Y \subseteq I$, we say that $X$ \textbf{merit-dominates} $Y$, written $X \succeq^{M} Y$, if $|X| \ge |Y|$ and, for every $k \le |Y|$, the $k$th highest merit-score individual of $X$ has a weakly higher merit score than the $k$th highest merit-score individual of $Y$. 
\end{quotation}

We say that $X$ \emph{strictly merit-dominates} $Y$, written $X \succ^M Y$, if $X \succeq^M Y$ and either $|X| > |Y|$, or there exists $k \in \{1, \dots, |Y|\}$ such that the $k$th highest merit-score individual of $X$ has a strictly higher merit score than the $k$th highest merit-score individual of $Y$.

The merit-based comparison of two sets can be extended to compare choice rules.

\begin{definition}
	A choice rule $C$ \textbf{merit-dominates }another choice rule $C^{'}$, $C \succeq^{M} C^{'}$, if, for all sets of individuals $A\subseteq I$, $C\left(A\right)$ merit-dominates $C^{'}(A)$. A choice rule $C$ \emph{strictly merit-dominates} another choice rule $C'$, written $C \succ^M C'$, if $C \succeq^M C'$ and there exists a set of applicants $A \subseteq I$ such that $C(A) \succ^M C'(A)$.
\end{definition}

We say that an assignment rule $\eta$ \textbf{merit-dominates} another assignment rule $\eta^{'}$, written $\eta \succeq^{M} \eta^{'}$, if $C^{\eta} \succeq^{M} C^{\eta^{'}}$. An assignment rule $\eta$ \emph{strictly merit-dominates} another assignment rule $\eta'$, written $\eta \succ^M \eta'$, if $C^{\eta} \succ^M C^{\eta'}$.

\section{Indra Sawhney (1992) Principles as Axioms}\label{sec:IS}

Indra Sawhney (1992) defines SC, ST, and OBC reservations as \textbf{hard} constraints and formulates three desiderata on the selection procedure: the \textbf{over-and-above (OA) principle}, \textbf{within-category fairness (WCF)}, and \textbf{quota-filling subject to eligibility (QFE)}. We introduce them in turn.

The first requires that open-category positions be allocated to the highest-scoring applicants up to the open-category's capacity. In doing so, positions reserved for reserved categories can be more effectively assigned to individuals who would otherwise not qualify for them without the reservation policy.

\begin{definition}\label{def:OA}
   An assignment rule $\eta$ satisfies the \textbf{over-and-above (OA) principle} if, for all $(A,T) \in 2^{I} \times \mathcal{T}$,  $$\mid\left\{ j | \left(j,o\right)\in\eta(A,T)\right\}\mid =\min\left\{ |A|,q^{o}\right\},$$ and  for all $i\in A$ such that $rank_{A}(i) \leq q^{o}$, $$(i,o)\in \eta(A,T).$$
\end{definition}

The first part of the OA principle states that if there are fewer than $q^{o}$ applicants, they are all accepted through the open-category. Moreover, it emphasizes the fact that reserves are considered as \textit{hard} reserves in the sense that exactly $q^{o}$ open-category positions are assigned whenever there are at least $q^{o}$ applicants. The second part states that the top $q^{o}$ applicants must be assigned open-category positions.

The next is a fairness condition that requires that merit be respected both across categories in the allocation of open-category positions and within each category in the allocation of reserved positions. If an applicant is assigned an open-category position, every higher-merit applicant, regardless of category membership, must also receive a position. If an applicant is assigned a reserved position, every higher-merit applicant belonging to the same reserve category must also receive a position. In either case, the condition does not prescribe the type of position assigned to the higher-merit applicant; it only requires that the applicant not be left unassigned.

\begin{definition}\label{def:WCF}
An assignment rule $\eta$ satisfies \textbf{within-category fairness} (WCF) if, for any pair of individuals $i,j\in A$ such that $i\succ j$ and $(j,\tau)\in \eta(A,T) \text{, for some } \tau\in \{o\}\cup R$, the following conditions hold. 
\[
\begin{aligned}
\tau=o
&\implies (i,\tau')\in\eta(A,T),
\quad \text{for some } \tau'\in\{o,t(i)\},\\
\tau\in R,\ t(i)=t(j)
&\implies (i,\tau')\in\eta(A,T),
\quad \text{for some } \tau'\in\{o,t(i)\}.
\end{aligned}
\]
\end{definition}

The last is a mild efficiency property. It states that for each reserve category, positions can remain unfilled only if no eligible applicant is left unassigned.

\begin{definition}\label{def:QFE}
An assignment rule $\eta$ satisfies \textbf{quota-filling subject to eligibility} (QFE) if, for any $r\in R$, whenever an individual $i$ with $t(i)=r$ is not assigned, $$|\{i\in A: (i,r)\in \eta(A,T)\}|=q^{r}.$$
\end{definition}

The judgment in Indra Sawhney (1992) proposes the following simple two-stage choice procedure, which we call the \textit{Indra Sawhney Assignment Rule}.

\subsection{The Indra Sawhney Assignment Rule}
Given a set of applicants $A\subseteq I$ and a profile of category membership $T=\left(t(i)\right)_{i\in A}$, the assignment of applicants to categories is calculated as follows:

\paragraph{\textbf{Stage 1}}
Only open-category positions are considered. For each applicant $i\in A$ such that $rank_{A}(i)\leq q^{o}$, add $(i,o)$ to $\eta^{o}(A,T)$. Let $C^{o}(A)$ be the set of individuals chosen. 

\paragraph{\textbf{Stage 2}}
Among the remaining applicants $A^{1}=A\setminus C^{o}\left(A\right)$, for each reserve category $r\in R$, applicants are chosen one at a time according to $\succ^{r}$ up to capacity $q^{r}$.  Let us call the set of applicants chosen for category $r$ by $C^{r}\left(A^{1}\right)$. For each individual $i$ chosen in category $r$, add $(i,r)$ to $\eta^{r}(A^{1},T)$.

Then $$\eta^{IS}(A,T)=\eta^{o}\left(A,T\right)\cup\underset{r\in R}{\bigcup} \eta^{r}\left(A^{1},T\right).$$

The three conditions above characterize the Indra Sawhney assignment rule, which has been used in India for decades when the reserves in each vertical category are \textit{hard} reserves. The following result, Theorem \ref{thm:IS}, is the starting point of our axiomatic analysis.\footnote{An indirect (in plain English) characterization of the same choice rule is given in Section 2.2 of \cite{sonmez2022affirmative}, in which the OA principle is defined \textbf{differently} as follows: ``\textit{Next, VR protected positions must be allocated on an ``over-and-above" basis; that is, positions that can be received without invoking the VR protections do not count against VR-protected positions}." Note that their definition does not require that the highest-scoring applicants be assigned to the open-category. The authors do not provide a formal proof for their characterization claim.} 

\setcounter{theorem}{-1} 
\begin{theorem}\label{thm:IS}
An assignment rule satisfies the OA principle, WCF, and QFE if and only if it is $\eta^{IS}$.
\end{theorem}

\section{Ashoka Kumar Thakur (2008): Two Interpretations of the OBC Cutoff}\label{sec:AKT}
 The October 14, 2008, clarification to the judgment in \textit{Ashoka Kumar Thakur} (2008) adds an eligibility restriction to the \textit{Indra Sawhney} framework. It stated that OBC ``cutoff marks'' could be relaxed by at most 10\% relative to the open-category cutoff. However, the legal meaning of ``cutoff'' remained ambiguous.

Under \textbf{endogenous interpretation}, the relevant open-category cutoff is the score of the last applicant assigned an open-category position. OBC eligibility therefore depends on the realized assignment and is endogenous to the applicant pool. Under the \textbf {exogenous interpretation}, the relevant cutoff is the minimum eligibility or qualifying mark announced \textit{before admissions begin}, so OBC eligibility is exogenous to the allocation. JNU implemented the first interpretation in 2008--09 and 2009--10.

We denote by $\lambda \geq 0$ the maximum cutoff score difference allowed between the realized open-category cutoff and the OBC category. An OBC applicant is eligible for an OBC position only if her score is no more than $\lambda$ points below the open-category cutoff score, that is, the score of the lowest-scoring applicant assigned an open-category position. We formulate the \textbf{endogenous} cutoff interpretation as a formal axiom in the following. 

\begin{definition}\label{def:mQFE}
    An assignment rule $\eta$ satisfies the \textit{modified quota-filling subject to eligibility} (\textbf{mQFE}) if for any set of applicants $A$ and $T=(t(i))_{i\in A}$, the following holds.
    
\begin{enumerate}
        \item If an individual $i$ with $t(i)=r \in R \setminus \{OBC\}$ is not assigned, then the number of individuals who are matched to a category $r$ position must be equal to $q^{r}$. 
        
        \item If an individual $i$ with $t(i)=OBC$ is not assigned and $$\theta (i) \geq \min\{\theta(j)|(j,o)\in \eta(A,T)\}-\lambda,$$ then the number of individuals assigned OBC positions must be $q^{OBC}$.
        
        \item If an individual $i$ with $t(i)=OBC$ and $$\theta (i) < \min\{\theta(j)|(j,o)\in \eta(A,T)\}-\lambda,$$ then $(i,OBC)\notin \eta(A,T)$.
\end{enumerate}
          
\end{definition}

Conditions (1) and (2) are the QFE requirement applied, respectively, to categories other than OBC and to OBC applicants who clear the cutoff. Condition (3) states that OBC applicants whose scores fall more than $\lambda$ points below the open-category cutoff cannot be assigned OBC positions.

\subsection{Endogenous Cutoff without De-reservation}

A simple modification of the Indra Sawhney assignment rule straightforwardly implements the mQFE principle in conjunction with the OA principle and WCF in the absence of OBC de-reservation.

Given a set of applicants $A \subseteq I$, the profile of category membership $T=\left(t(i)\right)_{i\in A}$, their test scores $(\theta (i))_{i\in A}$, and the maximum cutoff score difference $\lambda$, the assignment rule $\eta^{AKT}$ is calculated as follows:

\paragraph{\textbf{Stage 1}}
Only open-category positions are considered. For each applicant $i\in A$ such that $rank_{A}(i)\leq q^{o}$, add $(i,o)$ to $\eta^{o}(A,T)$. Let $C^{o}(A)$ be the set of individuals chosen. 

\paragraph{\textbf{Stage 2}}
Among the remaining applicants $A^{1}=A\setminus C^{o}\left(A\right)$, for reserve categories $r\in R \setminus \{OBC\}$, the applicants are chosen one at a time following $\succ^{r}$ up to capacity $q^{r}$. Call the set of applicants chosen for the reserved category $r$ by
$C^{r}\left(A^{1}\right)$. For each individual $i$ chosen in category $r\in R \setminus \{OBC\}$, add $(i,r)$ to $\eta^{r}(A^{1},T)$.

\paragraph{\textbf{Stage 3}}
Among the remaining applicants, consider those in the OBC category. Let $j^{*}$ be the lowest-scoring applicant assigned an open-category position in Stage 1. For each individual $i$ with $t(i)=OBC$ and $\theta (i) < \theta(j^{*})-\lambda$, $(i,OBC) \notin \eta^{OBC}(A^{1},T)$. Among the applicants in $\{i\in A^{1}| t(i)=OBC \text{  and  } \theta (i) \geq \theta(j^{*})-\lambda\}$, select the highest-scoring $q^{OBC}$. If there are fewer than $q^{OBC}$ such applicants, select all of them. For each chosen $i$, add $(i,OBC)$ to $\eta^{OBC}(A^{1},T)$. 

Then $$\eta^{AKT}(A,T)=\eta^{o}\left(A,T\right) \cup \underset{r\in R}{\bigcup} \eta^{r}\left(A^{1},T\right).$$

The sole distinction between the assignment rule $\eta^{AKT}$ and the Indra Sawhney assignment rule $\eta^{IS}$ is that $\eta^{AKT}$ restricts the eligibility of OBC seats to individuals whose score meets or exceeds the open-category cutoff minus $\lambda$. We characterize the assignment rule $\eta^{AKT}$ as follows.

\begin{theorem}\label{thm:AKT}
    An assignment rule satisfies the OA principle, WCF, and mQFE if and only if it is $\eta^{AKT}$.
\end{theorem}

Let $C^{AKT}$ be the choice rule induced by the assignment rule $\eta^{AKT}$, which is a straightforward extension of the Indra Sawhney choice rule $C^{IS}$ and satisfies conditions that have been key in the market design literature.

\begin{proposition}\label{prop:JNU}
    The choice rule $C^{AKT}$ is consistent and satisfies the substitutes property. 
\end{proposition}

\subsection{The Waste Generated by the Cutoff Mandate in the Absence of De-reservation}\label{sec:AKTwaste}

Although Theorem \ref{thm:AKT} shows that the cutoff mandate entails a unique assignment rule, that rule has a serious flaw. It can be very wasteful. Every individual assigned a position under $\eta^{AKT}$ receives an assignment under $\eta^{IS}$, but the converse does not hold. OBC applicants who do not reach the elevated eligibility threshold are assigned in $\eta^{IS}$ but not in $\eta^{AKT}$. 

\begin{remark}\label{rem:waste}
For every set of applicants $A\subseteq I$, $C^{AKT}(A)\subseteq C^{IS}(A)$. The difference $C^{IS}(A)\setminus C^{AKT}(A)$ consists of OBC applicants who are assigned OBC positions under $\eta^{IS}$ but are ineligible under the cutoff mandate; the corresponding OBC positions remain vacant under $\eta^{AKT}$.
\end{remark}

The normatively troubling feature of $\eta^{AKT}$ relative to $\eta^{IS}$ is the \textbf{waste} it generates. Positions that could be filled by eligible applicants under the Indra Sawhney rule are simply left vacant.\footnote{\cite{baswanaetal2019} documents a substantial number of vacancies in the OBC category for admissions to engineering colleges. OBC de-reservation was one of the stated "business rules" in the redesign of engineering college admissions mechanism. Moreover, a cursory examination of online resources reveals the widespread prevalence of unfilled OBC positions throughout India.} OBC vacancies are already widespread \textit{without} any cutoff restriction. Thus, conditioning OBC eligibility on the open-category cutoff further aggravates these vacancies.\footnote{A similar cutoff score restriction exists in admissions to high schools in China (\cite{dur2024gets} and \cite{hu2025verifiable}). Wuhan eliminated the 30-point cutoff score constraint for reserved seats in 2024. \cite{hu2025verifiable} argues that China is trending toward completely eliminating this constraint.} 

The choice rule $C^{AKT}$ also violates size monotonicity. The following example illustrates these problems.

\begin{example}\label{ex:waste}
Consider an institution with three positions, one of which
is open-category and two of which are reserved for OBC.
That is, $(q^o,q^{OBC})=(1,2)$ and $q^r=0$ for all
$r\in R\setminus\{OBC\}$. Let $\lambda=10$.
Consider the set of individuals $I=\{i_1,i_2,i_3,i_4\}$
with the following merit scores and category memberships:

\begin{center}
\begin{tabular}{ccc}
\hline
Individual & Category & Score \\
\hline
$i_1$ & $g$   & $100$ \\
$i_2$ & $g$   & $90$  \\
$i_3$ & $OBC$ & $85$  \\
$i_4$ & $OBC$ & $84$  \\
\hline
\end{tabular}
\end{center}

Let $B=\{i_2,i_3,i_4\}$. Under $\eta^{AKT}$,
individual $i_2$ is assigned the open-category position.
The open-category cutoff is $\theta(i_2)=90$, and hence
the OBC eligibility threshold is $90-10=80$.
Since both $i_3$ and $i_4$ satisfy the eligibility
requirement, they are assigned OBC positions. Thus,
\[
\eta^{AKT}(B,T)
=
\{(i_2,o),(i_3,OBC),(i_4,OBC)\}.
\]
The induced choice is
\[
C^{AKT}(B)=\{i_2,i_3,i_4\}.
\]
All three positions are filled.

Now consider the enlarged applicant set
$A=B\cup\{i_1\}=I$. Under $\eta^{AKT}$, individual
$i_1$ is assigned the open-category position, raising
the open-category cutoff to $\theta(i_1)=100$.
Consequently, the OBC eligibility threshold increases
to $100-10=90$. Neither $i_3$ nor $i_4$ satisfies
the new eligibility requirement, and both OBC positions
remain vacant. Thus,
\[
\eta^{AKT}(A,T)=\{(i_1,o)\},
\]
and the induced choice is
\[
C^{AKT}(A)=\{i_1\}.
\]

Since $B\subsetneq A$ but
\[
|C^{AKT}(B)|=3>1=|C^{AKT}(A)|,
\]
the choice rule $C^{AKT}$ violates size monotonicity.

Moreover, under the Indra Sawhney assignment rule,
we have
\[
\eta^{IS}(A,T)
=
\{(i_1,o),(i_3,OBC),(i_4,OBC)\}.
\]
Hence,
\[
C^{AKT}(A)=\{i_1\}
\subsetneq
\{i_1,i_3,i_4\}=C^{IS}(A).
\]
The endogenous cutoff requirement therefore generates
two vacant OBC positions that would otherwise be filled
under $\eta^{IS}$.
\end{example}

Although $C^{AKT}$ satisfies path independence, it can be wasteful. In Example \ref{ex:waste}, two seats are wasted as $C^{AKT}(I)=\{i_1\}$. If the number of OBC candidates whose scores exceed the cutoff is low, the waste can be substantial. This outcome highlights the critical necessity of the directive of the 2008 correction to Ashoka Kumar Thakur (2008) on the reversion of unfilled OBC seats, to which we next turn.

\section{Endogenous Cutoffs}\label{sec:endcutoff}

Under the endogenous cutoff interpretation, the eligibility threshold for OBC applicants is determined by \textbf{the score of the last applicant} admitted to an open-category position. When unfilled OBC positions are de-reserved, this interpretation admits two alternative implementations, \textit{depending on whether the transferred positions affect the open-category cutoff}. 

Under the first interpretation, the cutoff is determined by the last applicant admitted to an open-category position, including positions created through de-reservation. We establish an impossibility result: no assignment rule can simultaneously satisfy the OA principle, WCF, mQFE, and OBC de-reservation. Under the second interpretation, adopted by JNU, the cutoff is determined exclusively by the last applicant admitted to an original open-category position, excluding positions created through de-reservation. Although this interpretation avoids the impossibility and admits a merit-optimal assignment rule, the induced choice rule violates the substitutes condition. Thus, the two implementations of endogenous eligibility cutoffs confront distinct theoretical difficulties. One generates an incompatibility among fundamental allocation requirements, while the other produces complementarities in institutional choice. These findings provide an economic rationale for the exogenous cutoff interpretation established in P.V. Indiresan (2011), under which eligibility is determined in advance and the admissible rules with de-reservation satisfy substitutability.

\subsection{Expanded-open cutoff: An impossibility}
We first analyze the case in which the OBC cutoff is determined from the open-category cutoff according to the last applicant admitted to an open-category position \textbf{overall, including any reverted OBC positions}. Each newly created open position can lower the realized open-category cutoff and thereby enlarge the set of OBC applicants who are eligible. We show that this feedback can make the endogenous cutoff requirement, OA, WCF, and de-reservation jointly impossible. 

Under the endogenous cutoff interpretation, de-reserved positions are assigned under the label $o$, so more than $q^{o}$ positions may be allocated as open-category.\footnote{This is how OBC de-reservation has been implemented in engineering admissions (\cite{baswanaetal2019}). The reversion is made through rerunning the deferred acceptance algorithm on all applicants by just adjusting the open category and OBC positions.} Accordingly, the OA principle needs a minor change: the equality in condition (i) of Definition \ref{def:OA} is replaced with ``$\geq$" as follows. 

 \begin{definition}\label{def:softOA}
   An assignment rule $\eta$ satisfies the \textbf{over-and-above (OA) principle} (soft-reserve version) if, for all $(A,T) \in 2^{I} \times \mathcal{T}$, $$\mid\left\{ j | \left(j,o\right)\in\eta(A,T)\right\}\mid \geq \min\left\{ |A|,q^{o}\right\},$$ and, for all $i\in A$ such that $rank_{A}(i) \leq q^{o}$, $$(i,o)\in \eta(A,T).$$
\end{definition}

We formalize the de-reservation requirement as follows. Whenever positions in the combined open and OBC categories remain vacant, no applicant can be left unassigned.

\begin{definition}\label{def:dereservation}
    An assignment rule $\eta$ satisfies \textbf{OBC de-reservation} if for any set of individuals $A$ and $T=(t(i))_{i\in A}$, 
    $$|\{(i,\tau)\in \eta(A,T): \tau \in \{o,OBC\}\}|<q^{o}+q^{OBC} \implies |\eta(A,T)|=|A|.$$
\end{definition}

When reverted OBC positions are relabeled as open-category and filled, the open-category cutoff score becomes lower. In turn, some OBC candidates who were ineligible due to their low scores may become eligible under the lower open-category cutoff. This feedback between de-reservation and eligibility makes the joint implementation of the cutoff mandate and the de-reservation policy impossible.

\begin{theorem}\label{thm:impossibility}
    No assignment rule satisfies the OA principle (soft reserves version), WCF, OBC de-reservation, and mQFE.
\end{theorem}

\begin{proof}
Consider an institution with $q=2$ positions, of which one is of open-category and one is reserved for OBC. That is, $(q^o,q^{OBC})=(1,1)$ and $q^r=0$ for all
$r\in R\setminus\{OBC\}$. Let $\lambda=10$ and consider the set of applicants
$I=\{i_1,i_2,i_3\}$ with the following merit scores and category memberships:

\begin{center}
\begin{tabular}{ccc}
\hline
Applicant & Category & Merit score \\
\hline
$i_1$ & $g$ & $100$ \\
$i_2$ & $g$ & $90$ \\
$i_3$ & $OBC$ & $89$ \\
\hline
\end{tabular}
\end{center}

Suppose, toward a contradiction, that an assignment rule $\eta$ satisfies the four axioms. By the OA principle, $(i_1,o)\in\eta(I,T)$. Given the category memberships of the remaining applicants and the capacity constraints, exactly one
of the following four assignments must arise.

\begin{enumerate}
    \item $\eta(I,T)=\{(i_1,o),(i_3,OBC)\}$.
    The open-category cutoff is $100$, and hence
    the OBC eligibility threshold is $90$.
    Since $\theta(i_3)=89<90$, assigning $i_3$
    an OBC position violates part (3) of mQFE.

    \item $\eta(I,T)=\{(i_1,o),(i_2,o)\}$.
    Under the expanded-open cutoff interpretation,
    the open-category cutoff is $90$, and hence
    the OBC eligibility threshold is $80$.
    Since $\theta(i_3)=89\geq80$, applicant $i_3$
    is eligible for an OBC position but remains
    unassigned. Moreover, no applicant is assigned
    an OBC position, although $q^{OBC}=1$.
    This violates part (2) of mQFE.

    \item $\eta(I,T)=\{(i_1,o),(i_3,o)\}$.
    Since $i_2\succ i_3$ and $(i_3,o)\in\eta(I,T)$,
    WCF requires that $i_2$ also be assigned a
    position, contradicting WCF.

    \item $\eta(I,T)=\{(i_1,o)\}$.
    Since
    \[
    |\{(i,\tau)\in\eta(I,T):
    \tau\in\{o,OBC\}\}|
    =1<q^o+q^{OBC}=2,
    \]
    OBC de-reservation requires that
    $|\eta(I,T)|=|I|$.
    However, $|\eta(I,T)|=1<3=|I|$,
    contradicting OBC de-reservation.
\end{enumerate}

Since every feasible assignment satisfying the OA principle violates at least one of the remaining three axioms, no assignment rule satisfies all axioms simultaneously.
\end{proof}

This impossibility is specific to an endogenous cutoff interpretation when reverted vacant OBC positions are allowed to alter the open-category cutoff. According to the JNU interpretation, this impossibility disappears.

\subsection{Original-open cutoff: The JNU Interpretation} \label{sec:etaJNU}

The assignment rule used by JNU, $\eta^{JNU}$, keeps the open-category cutoff fixed at the level determined by the original $q^{o}$ open-category positions. That is, de-reserved positions do not enter the computation of the cutoff. Operationally, the endogenous cutoff is set first from the original open-category positions. The OBC positions that remain unfilled under that cutoff then revert and are allocated to the highest-scoring remaining applicants. 

A natural implementation of this reading is to assign de-reserved OBC positions a distinct label ``$d$", reserving the label ``$o$" exclusively for original open-category positions, and to determine the OBC eligibility cutoff solely by reference to the lowest-scoring applicant admitted through an ``$o$" position.

The only design question the opinion leaves open is \textit{when}, in the precedence order, the reverted positions should be filled---before or after the other reserve categories are processed. The JNU assignment rule filled vacant OBC positions after processing the reserve category seats with the highest-scoring members of the remaining applicants. 

Formulating this idea requires a minor extension of the model. Let $D$ be the new category representing de-reserved OBC positions. We say that an individual $i$ is assigned a position in category $D$ under an assignment rule $\eta$ if $(i,d) \in \eta(A,T)$. With this change, the definition of an assignment rule is updated as follows.

\begin{definition}\label{def:assignmentruleD}
An \textbf{assignment rule}  $\eta$ is a function in which, given a set of applicants $A$ and a category membership profile $T$, $$\eta(A,T) = \{(i,\tau)| i\in A  \text{ and } \tau \in \{o,d\}\cup R\} \text{ such that }$$
 
	\begin{enumerate}
        \item [(i)] for any $i\in A$, we have at most one $(i,\tau)\in \eta(A,T)$, where
		\[
		\begin{cases}
			\begin{array}{c}
				\tau \in \{o,d\} \\
				\tau \in\left\{ t(i),o,d\right\}  
			\end{array} & \begin{array}{c}
				\text{if} \;t(i)=g,\\
				\text{if} \;t(i)\in R,
		\end{array}\end{cases}
		\]
		\item [(ii)] $\lvert \eta(A,T) \lvert \leq \overline{q}$, and
        \item [(iii)] $\mid\left\{ (i,\tau)\in \eta(A,T): \tau=r \right\} \mid\leq q^{r}$, for each $r\in R$.
        \item[(iv)] $\left|\left\{(i,\tau)\in\eta(A,T): \tau\in\{d,OBC\}\right\}\right| \leq q^{OBC}$.
	\end{enumerate}
\end{definition}

With the introduction of category-$d$ positions,
the choice rule induced by an assignment rule $\eta$
is defined as
\[
C^\eta(A)
=
\left\{
i\in A:
(i,\tau)\in\eta(A,T)
\text{ for some }\tau\in\{o,d\}\cup R
\right\}.
\]

\paragraph{\textbf{The JNU Assignment Rule}}

Given a set of applicants $A$, the profile $T=(t(i))_{i\in A}$, their test scores $(\theta (i))_{i\in A}$, and the maximum cutoff score difference $\lambda$, the assignment of applicants to categories is calculated as follows:

\paragraph{\textbf{Stage 1}}
Only open-category positions are considered. For each applicant $i\in A$ such that $rank_{A}(i)\leq q^{o}$, add $(i,o)$ to $\eta^{o}(A,T)$. Let $C^{o}(A)$ be the set of individuals chosen. 

\paragraph{\textbf{Stage 2}}
Among the remaining applicants $A^{1}=A\setminus C^{o}\left(A\right)$, for reserve categories $r\in R \setminus \{OBC\}$, the applicants are chosen one at a time according to $\succ^{r}$ up to capacity $q^{r}$. Call the set of applicants chosen for the reserved category $r$ by
$C^{r}\left(A^{1}\right)$. For each individual $i$ chosen in category $r$, add $(i,r)$ to $\eta^{r}(A^{1},T)$.

\paragraph{\textbf{Stage 3}}
Among the remaining applicants, consider those in the OBC category. Let $j^{*}$ be the lowest-scoring applicant assigned an original open-category position in Stage 1, so that the open-category cutoff score is $\theta(j^{*})$. An OBC applicant $i$ is \textit{eligible} if $\theta (i) \geq \theta(j^{*})-\lambda$. Among the remaining eligible OBC applicants, select the highest-scoring $q^{OBC}$. If the number of such applicants is less than $q^{OBC}$, then choose all of them. For each chosen $i$, add $(i,OBC)$ to $\eta^{OBC}(A^{1},T)$. Let $\kappa$ be the number of vacant OBC positions at the end of this stage.

\paragraph{\textbf{Stage 4}}
Consider the remaining applicants $A^{''}$. Among them, select the highest-scoring $\kappa$ applicants, regardless of category membership. If the number of remaining applicants is less than $\kappa$, select all of them. For each individual $i$ chosen in this stage, add $(i,d)$ to  $\eta^{D}(A^{''},T)$, where $d$ denotes a de-reserved OBC position. Then,

 $$\eta^{JNU}(A,T)=\eta^{o}\left(A,T\right)\cup\underset{r\in R}{\bigcup} \eta^{r}\left(A^{1},T \right) \cup \eta^{D}(A^{''},T).$$ 

The assignment rule $\eta^{JNU}$ runs $\eta^{AKT}$ in its first three stages. Unlike $\eta^{AKT}$, it utilizes all vacant OBC positions in its final stage and fills them based on merit scores as category-$D$ positions. 

Let $C^{JNU}$ be the choice rule induced by the assignment rule $\eta^{JNU}$. The following example illustrates how $\eta^{JNU}$ eliminates the waste generated by the cutoff mandate.

\begin{example}\label{ex:etaJNU}
    Consider an institution with $\overline{q}=7$. Let the distribution of positions over categories be $(q^{o},q^{SC},q^{ST},q^{OBC},q^{EWS})=(3,1,1,2,0)$. Let $\lambda =10$. Suppose that $I=\left\{ i_{1},...,i_8\right\}$ with the following test scores and category memberships. 

\[
\begin{array}{c|c|c}
\text{Applicant} & \text{Category} & \text{Score}\\
\hline
i_{1} & g & 100\\
i_{2} & g & 99\\
i_{3} & SC & 97\\
i_{4} & g & 96\\
i_{5} & g & 90\\
i_{6} & OBC & 85\\
i_{7} & ST & 84\\
i_{8} & SC & 83
\end{array}
\]

We first compute $\eta^{AKT}(I,T)$. In Stage 1, the top three applicants $i_1$, $i_2$, and $i_3$ are assigned open-category positions, so the open-category cutoff is $\theta(i_3)=97$. In Stage 2, $i_8$ and $i_7$ are assigned the SC and ST positions, respectively. In Stage 3, the only OBC applicant, $i_6$, has score $85<97-10=87$ and is therefore ineligible. Both OBC positions remain vacant:
$$\eta^{AKT}(I,T)=\{(i_1,o),(i_2,o),(i_3,o),(i_8,SC),(i_7,ST)\}.$$
Two of the seven positions are wasted, while the high-scoring applicants $i_4$ and $i_5$ are rejected.

The rule $\eta^{JNU}$ coincides with $\eta^{AKT}$ in its first three stages and then, in Stage 4, fills the $\kappa=2$ vacant OBC positions with the highest-scoring remaining applicants, $i_4$ and $i_5$:
$$\eta^{JNU}(I,T)=\{(i_1,o),(i_2,o),(i_3,o),(i_8,SC),(i_7,ST),(i_4,d),(i_5,d)\}.$$
The induced choice is $C^{JNU}(I)=\{i_1,i_2,i_3,i_4,i_5,i_7,i_8\}$. All seven positions are filled, and they are filled by the most meritorious applicants consistent with the reservation policy and the cutoff mandate.
\end{example}

The assignment rule $\eta^{JNU}$ is the merit-optimal implementation that would arise if the Court had retained JNU's endogenous interpretation while also requiring de-reservation of residual OBC vacancies. Our next results show that the desirable performance of $\eta^{JNU}$ in Example \ref{ex:etaJNU} is not a coincidence, but is, in general, true. 

\subsection{Characterization}
We next show that $\eta^{JNU}$ is the merit-optimal de-reservation rule within the JNU interpretation. To do so, we first restate our axioms with the relabeling of de-reserved OBC positions as category-$D$ positions.

\begin{definition}\label{def:OAstar}
An assignment rule $\eta$ satisfies the \textbf{over-and-above$^{*}$ ($OA^{*}$) principle} if, for every $A\subseteq I$ and $T\in\mathcal{T}$, the following conditions hold:
\begin{enumerate}
    \item The number of applicants assigned to open-category positions is
    $\min\{|A|,q^o\}$. That is,
    \[
    \left|\{i\in A:(i,o)\in\eta(A,T)\}\right|
    =\min\{|A|,q^o\}.
    \]

    \item Every applicant assigned to an open-category position has higher merit than any applicant assigned to a category-$d$ or reserved-category position. That is, for all $i,j\in A$ and
    $\tau\in\{d\}\cup R$,
    \[
    (i,o)\in\eta(A,T)
    \text{ and }
    (j,\tau)\in\eta(A,T)
    \implies i\succ j.
    \]
\end{enumerate}
\end{definition}

We now modify the WCF property in the presence of category-d positions.

\begin{definition}\label{def:WCFstar}
An assignment rule $\eta$ satisfies \textbf{within-category fairness$^{*}$} (WCF$^{*}$) if, for any pair of individuals $i,j\in A$ such that $i\succ j$ and $(j,\tau)\in \eta(A,T) \text{, for some } \tau\in \{o,d\}\cup R$, the following conditions hold. 
\[
\begin{aligned}
\tau \in \{o,d\}
&\implies (i,\tau')\in\eta(A,T),
\quad \text{for some } \tau'\in\{o,d,t(i)\},\\
\tau\in R,\ t(i)=t(j)
&\implies (i,\tau')\in\eta(A,T),
\quad \text{for some } \tau'\in\{o,d,t(i)\}.
\end{aligned}
\]
\end{definition}

The mQFE property remains the same. Note that since the label $o$ is now reserved exclusively for the original open-category positions, the cutoff $\min\{\theta(j)|(j,o)\in \eta(A,T)\}$ appearing in its statement is automatically determined by the original open-category positions only. 

Finally, the OBC de-reservation property reflects that unfilled OBC seats are allocated as category-$D$ positions: whenever positions in the combined OBC and de-reserved categories remain vacant, no applicant can be left unassigned.

\begin{definition}\label{def:dereservationstar}
    An assignment rule $\eta$ satisfies \textbf{OBC de-reservation$^{*}$} if for any set of individuals $A$ and $T=(t(i))_{i\in A}$, 
    $$|\{(i,\tau)\in \eta(A,T): \tau \in \{d,OBC\}\}|<q^{OBC} \implies |\eta(A,T)|=|A|.$$
\end{definition}

Let $\mathcal{E}$ be a class of assignment rules. We say that an assignment rule
$\eta \in \mathcal{E}$ is \emph{merit-optimal in} $\mathcal{E}$ if
$\eta \succeq^M \eta'$ for every $\eta' \in \mathcal{E}$.

Let $\mathbb{A}^*$ denote the class of assignment rules that satisfy the
OA$^*$ principle, WCF$^*$, mQFE, and OBC de-reservation$^*$.

Theorem \ref{thm:etaJNU} below confirms that JNU's choice of filling vacant OBC positions at the end based on merit scores is the optimal one if the aim is to maximize the meritocracy of the selected set of applicants. The resulting rule merit-dominates every rule consistent with the endogenous cutoff and the de-reservation requirement. 

\begin{theorem}\label{thm:etaJNU}
The assignment rule $\eta^{JNU}$ belongs to $\mathbb{A}^*$ and is
merit-optimal in $\mathbb{A}^*$. Moreover, every merit-optimal rule in
$\mathbb{A}^*$ induces the choice rule $C^{JNU}$.
\end{theorem}

Theorem \ref{thm:etaJNU} establishes that $\eta^{JNU}$---the rule that fills the de-reserved positions last---is optimal with respect to merit within $\mathbb{A}^{*}$. 

Although the JNU assignment rule is merit-optimal among the assignment rules in $\mathbb{A}^{*}$, its induced choice rule \textbf{violates the substitutes condition}. Importantly, this failure is not caused by endogenous eligibility cutoffs alone. As Proposition 1 establishes, the AKT choice rule satisfies substitutability despite employing an endogenous cutoff. Rather, the violation arises from the interaction between endogenous eligibility and de-reservation. Adding an applicant may raise the open-category cutoff, render previously eligible OBC
applicants ineligible, and thereby increase the number of positions transferred to open competition. As a result, an applicant who was previously rejected may become selected. This complementarity disappears under the exogenous eligibility framework considered in Section 6, where every admissible transfer rule satisfies the substitutes condition.

\begin{proposition}\label{prop:etaJNU}
    The choice rule $C^{JNU}$ is consistent and size monotonic, but violates the substitutes condition.
\end{proposition}

Proposition \ref{prop:etaJNU} establishes that $\eta^{JNU}$ is not DA-compatible. To show the violation of the substitutes condition, we provide the following example. 

\begin{example}
Consider an institution with three positions, one of which is open-category, one is reserved for SC, and one is reserved for OBC. Thus,
\[
(q^o,q^{SC},q^{OBC})=(1,1,1),
\]
with all other reserved-category capacities equal to zero.
Let $\lambda=20$.

Consider four applicants with the following category memberships and merit scores:
\[
\begin{array}{c|c|c}
\text{Applicant} & \text{Category} & \text{Score}\\
\hline
i_1 & g   & 40\\
i_2 & SC  & 30\\
i_3 & g   & 20\\
i_4 & OBC & 10
\end{array}
\]

First, consider the applicant set $A=\{i_2,i_3,i_4\}$. Under $\eta^{JNU}$, applicant $i_2$ is assigned the original open-category position. 
The open-category cutoff becomes $30$, and the OBC eligibility threshold is $30-20=10$. Thus, applicant $i_4$ is eligible for and assigned
the OBC position. The SC position remains vacant and applicant $i_3$ is rejected. Thus, $\eta^{JNU}(A,T) =\{(i_2,o),(i_4,OBC)\}$ and $C^{JNU}(A)=\{i_2,i_4\}$.

Now consider the enlarged applicant set $B=A\cup\{i_1\}$. Under $\eta^{JNU}$, applicant $i_1$ is assigned the original open-category position, raising the
open-category cutoff to $40$. Applicant $i_2$ is then assigned the SC position. The OBC eligibility threshold is now $40-20=20$. Since $\theta(i_4)=10<20$, applicant $i_4$ is ineligible for the OBC position. Consequently, the vacant OBC position is de-reserved and converted into a category-$D$ position. Among the remaining applicants, $i_3$ has the highest merit score and is therefore assigned
the category-$D$ position. Hence, $\eta^{JNU}(B,T)=\{(i_1,o),(i_2,SC),(i_3,d)\}$ and $C^{JNU}(B)=\{i_1,i_2,i_3\}$.

It follows that $i_3\notin C^{JNU}(A)$ but $i_3\in C^{JNU}(A\cup\{i_1\})$. Thus, the choice rule $C^{JNU}$ induced by $\eta^{JNU}$ does not satisfy substitutability.
\end{example}

Endogenous eligibility cutoffs combined with de-reservation can generate complementarities, whereas endogenous cutoffs without de-reservation preserve substitutability under \(\eta^{AKT}\). De-reservation eliminates the waste associated with unfilled OBC positions, but its interaction with the eligibility rule determines whether the resulting choice rule preserves substitutability.

\begin{remark}\label{rem:JNU_cardinality}
For every set of applicants $A\subseteq I$, the cardinality
of the cohort selected under $\eta^{JNU}$ is given by
\[
|C^{JNU}(A)|
=
\min\left\{
|A|,
\min\{|A|,q^o\}
+
\sum_{r\in R\setminus\{OBC\}}
\min\{q^r,|A^{1,r}|\}
+
q^{OBC}
\right\},
\]
where $A^{1,r}=\{i\in A\setminus C^o(A):t(i)=r\}$.
In particular, the cardinality of the selected cohort is
independent of the maximum permissible cutoff score
difference $\lambda$.
\end{remark}

Remark~\ref{rem:JNU_cardinality} follows directly from
the proof of Proposition~2. Although changes in $\lambda$
may alter the composition of the selected cohort and the
types of positions assigned to individual applicants,
they do not affect the total number of positions filled.
Forward transfers completely neutralize the effect of
the endogenous OBC eligibility threshold on aggregate
position utilization. Thus, under $\eta^{JNU}$, the
cutoff mandate has distributional consequences but no
consequences for aggregate position utilization.

\section{P.V. Indiresan (2011): Exogenous Eligibility}\label{sec:adaptive}

\textit{P.V. Indiresan} (2011) resolved the ambiguity left by the 2008 litigation. The SCI held that the relevant ``cutoff marks'' are the minimum eligibility or qualifying marks announced in advance, not the score of the last applicant admitted in the open-category. The crucial implication for design purposes is that the eligibility set for OBC applicants is determined before assignment begins.

The Court also imposed a clear resolution on de-reservation. The OBC positions must first be offered, in merit order, to OBC applicants who satisfy the minimum eligibility score required for the OBC positions. Only if such applicants are unavailable should residual OBC positions be converted into open-category positions. In the language of our model, \textit{P.V. Indiresan} replaces the endogenous mQFE restriction with an exogenous eligibility constraint and then allows soft reserves only after the eligible OBC set has been exhausted.

The remaining design question is where the converted positions should appear in the precedence order. The Court specifies \textit{when} conversion becomes permissible, but it does not determine the relative position at which the resulting new open-category seats should be processed.

We introduce a family of sequential assignment rules that implement OBC de-reservation through iterative capacity transfers. The members of this family differ in how converted positions are distributed across the precedence order according to predetermined and cumulative transfer schemes. We establish that
transferring all vacant OBC positions forward yields the merit-optimal assignment rule within this family. Moreover, every member of the family induces a
choice rule that satisfies consistency, substitutability, and size monotonicity.

 \subsection{The Family of Iterative Capacity-Transfer
Assignment Rules}
An institution always fills the original open-category before the reserved categories. That is, the precedence ordering is 

\begin{equation}
    Open - c_1 - c_2 - c_3 - c_4,
\end{equation}

where $c_1,c_2,c_3,c_4 \in R$ such that $c_i \neq c_j$ for $i \neq j$. Let $\sigma=(c_1,c_2,c_3,c_4)$ denote a permutation of $R$. There are $4!=24$ such orderings. 

The OBC category considers only applicants whose score is greater than or equal to $\underline{\theta}^{OBC}$. If all $q^{OBC}$ positions are filled by such applicants, then no de-reservation is necessary. Otherwise, only the residual OBC vacancies are reverted to open-category. The point in the precedence order at which these additional open-category seats are filled has non-trivial distributional consequences. We expand the precedence order (1) to locate new groups of positions in the open categories $\{T_1,T_2,T_3, T_4,T_5\}$:

\begin{equation}
    Open - T_1 - c_1 - T_2 - c_2 - T_3 - c_3 - T_4 - c_4 - T_5.
\end{equation}

We call this precedence order the \textit{expanded precedence order}, and we refer to the groups $T_1,\ldots,T_5$ as the \textbf{de-reserved open-categories}. These are placeholders for positions originally reserved for OBC but reallocated as open-category positions, and the index $i$ of the group $T_i$ marks the point in the precedence order at which such positions can be filled. 

Each de-reserved open-category $T_i$ has an initial capacity of 0. When there are $\kappa$ unfilled positions in the OBC category, a vector $(r_1,r_2,r_3,r_4,r_5)$ with $\sum_{i=1}^{5}r_{i}=\kappa$ describes how these vacancies are distributed among the groups $T_1, T_2, T_3, T_4$, and $T_5$, where $r_i$ is the number of positions allotted to the group $T_i$, for all $i=1,...,5$.

Given the expanded precedence order (2), a set of individuals $A\subseteq I$, a category membership profile $T=(t(i))_{i\in A}$, the capacities of the groups of positions $\textbf{x}=(x_1,\dots,x_{10})\in \mathbb{Z}^{10}_{+}$, and scores of the applicants $(\theta(i))_{i\in A}$, the assignment rule $\eta^{\mathbf{x}}(A,T)$ induced by the capacity vector $\mathbf{x}$ is calculated as follows:

 \paragraph{\textbf{Step 1}}
Consider all applicants. Select the top-$x_1$ applicants according to the merit scores. If there are fewer than $x_1$ applicants, select all of them. Add $(i,o)$ to $\eta^{\mathbf{x}}(A,T)$ for any applicant $i\in A$ selected. Let $A^1$ be the set of remaining applicants.  

 \paragraph{\textbf{Step k}}[$2 \leq k \leq 10$]
Consider the \textit{relevant applicants}\footnote{The relevant applicants for the de-reserved open-categories $T_1$, $T_2$, $T_3, T_4$ and $T_5$ are \textbf{all} applicants in the set $A^{k-1}$. The relevant applicants for the SC, ST, and EWS categories are the remaining applicants who belong to the respective category. For the OBC category, the relevant applicants are the remaining applicants in $A^{OBC}$, that is, OBC applicants satisfying the predetermined minimum eligibility or qualifying threshold required by \textit{P.V. Indiresan}.} in $A^{k-1}$. The groups $\{T_i\}_{i=1}^{5}$ consider all applicants. Select top-$x_k$ applicants according to merit scores. If there are fewer than $x_{k}$ applicants, select all of them. For each applicant $i\in A^{k-1}$ selected in the groups $T_1,...,T_5$, add $(i,o)$ to $\eta^{\mathbf{x}}(A,T)$. Similarly, for each applicant $i$ who is selected in category $r\in R$, add $(i,r)$ to $\eta^{\mathbf{x}}(A,T)$. Let $A^{k}$ be the set of remaining applicants.

Next, we define a transfer scheme, where $\kappa \in \{0,...,q^{OBC}\}$ denotes the number of vacant positions in the OBC category.

\begin{definition}\label{def:transferscheme}
A \textit{transfer scheme} is a 5-dimensional vector function $f$ such that
\begin{enumerate}
\item $f(\kappa)=(f_{1}(\kappa),f_{2}(\kappa),f_{3}(\kappa),f_{4}(\kappa),f_{5}(\kappa))$ where $f_i(\kappa)\geq 0$ is the number of positions available in category $T_i$ and $\kappa=\sum_{j=1}^{5}f_{j}(\kappa)$,
 for each $\kappa \in \{0,\ldots,q^{OBC}\}$ . 
\item $f_{i}(x+1)-f_i(x)\in \{0,1\}$, for all $i\in \{1,2,3,4,5\}$ and $\sum_{i=1}^{5}[f_{i}(x+1)-f_i(x)]=1$, for all $x\in\{0,\dots,q^{OBC}-1\}.$
\end{enumerate}
\end{definition}

In other words, given $\kappa$, unfilled positions in the OBC category, a transfer scheme $f(\kappa)$ allocates these positions among the groups of positions $T_1, \dots,T_5$. The definition implies that each vacant position must be transferred to one of $T_1,...,T_5$.

We denote by $\mathcal{F}$ the set of all transfer schemes. The assignment rule induced by the transfer scheme $f\in \mathcal{F}$, $\eta^{f}$, is constructed as follows.

\subsection*{Assignment Rule Induced by Transfer Scheme f}
Set the initial capacity of each group $T_i$ to 0, $ \forall i=1,...,5$ and construct $$\mathbf{x^{1}}=(q^{o},0,q^{c_1},0,q^{c_2},0,q^{c_3},0,q^{c_4},0).$$ 

Let $x^{1}_{r^{*}}=q^{OBC}$ denote the initial capacity of the OBC category, where $r^{*}\in \{1,...,10\}$ denotes the position of the OBC seats in the precedence order. 

Given a set of applicants $A$, a category membership profile $T=(t(i))_{i\in A}$, and the vector of initial position distribution $\mathbf{q}=(q^{o},q^{SC},q^{ST},q^{OBC},q^{EWS})$, the assignment rule $\eta^{f}(A,T)$ selects the applicants in (possibly) multiple iterations as follows: 

\noindent\textbf{Iteration $n$ ($n \geq 1$).} Let $v_1 = 0$. Run $\eta^{x^n}(A,T)$ and let
$\kappa_n$ be the number of vacant OBC positions. Let $v_{n+1} = v_n + \kappa_n$ be the
cumulative number of vacant OBC positions, and define $x^{n+1}$ by
\[
  x^{n+1}_{2j} = f_j(v_{n+1}) \quad \text{for all } j = 1,\ldots,5,
  \qquad
  x^{n+1}_{r^*} = q^{OBC} - v_{n+1},
\]
with the capacities of the Open, SC, ST, and EWS categories unchanged.

The process ends when there are no vacant positions in the OBC category. Let $N$ be the last iteration. Then
$$
\eta^{f}(A,T)=\eta^{\mathbf{x^{N}}}(A,T).
$$

We denote by $\mathbb{A} \equiv\left(\eta^{f}\right)_{f\in\mathcal{F}}$ the set of assignment rules generated by different transfer schemes. Two notable assignment rules in $\mathbb{A}$ are the following. 

\paragraph{\textbf{Forward Transfer (FT) Assignment Rule}}
The FT assignment rule $\eta^{FT}$ is such that $f(\kappa)=(0,0,0,0,\kappa)$,
$\forall\kappa\in\left\{ 0,\ldots,q^{OBC}\right\} $. 

\paragraph{\textbf{Backward Transfer (BT) Assignment Rule}}
The BT assignment rule $\eta^{BT}$ is such that $f(\kappa)=(\kappa,0,0,0,0)$,
$\forall\kappa\in\left\{ 0,\ldots,q^{OBC}\right\} $.\footnote{The choice rule induced by the backward transfer assignment rule, $C^{BT}$, was introduced in \cite{aygun2022dereserve}.}

\begin{example}\label{ex:BTvFT}
Consider an institution with $\overline{q}=7$. The initial distribution of positions over categories is $(q^{o},q^{SC},q^{ST},q^{OBC},q^{EWS})=(3,1,1,2,0)$. Let $\underline{\theta}^{OBC}=80$. Suppose that $I=\left\{ i_{1},...,i_{8}\right\}$ with the following test scores and category memberships.

\[
\begin{array}{c|c|c}
\text{Applicant} & \text{Category} & \text{Score}\\
\hline
i_{1} & g & 100\\
i_{2} & g & 99\\
i_{3} & g & 98\\
i_{4} & SC & 97\\
i_{5} & g & 96\\
i_{6} & OBC & 85\\
i_{7} & ST & 84\\
i_{8} & SC & 83
\end{array}
\]

We first compute $\eta^{FT}(I,T)$. $i_{1}$, $i_{2}$, and $i_{3}$ are selected for open-category positions. $i_{4}$ and $i_{7}$ are assigned to the reserved SC and ST positions, respectively. $i_{6}$ is chosen for a reserved OBC position. The other OBC position remains unfilled and is made an open-category position. Among the unassigned individuals, $i_{5}$ has the highest score and is chosen for the open-category position. Thus, 
$$\eta^{FT}(I,T)=\{(i_{1},o),(i_{2},o),(i_{3},o),(i_{4},SC),(i_{5},o),(i_{6},OBC),(i_{7},ST)\}.$$

We now compute $\eta^{BT}(I,T)$. In the first iteration, $i_{1}$, $i_{2}$, and $i_{3}$ are selected in the open-category, $i_{4}$ is selected in the SC category, $i_{7}$ is selected in the ST category, and the only OBC individual $i_{6}$ is chosen in the OBC category. One reserved OBC seat remains vacant. Therefore, it becomes an open-category position by updating the capacity vector to $\left(4,1,1,1,0\right)$. 

In the second iteration, $i_{1}$, $i_{2}$ , $i_{3}$, and $i_{4}$ are selected in the open-category, $i_{8}$ is selected in the SC category, $i_{7}$ is selected in the ST category and $i_{6}$ is chosen in the OBC category. We end the procedure because there is no vacancy in the OBC category. Thus, we have
 $$\eta^{BT}(I,T)=\{(i_{1},o),(i_{2},o),(i_{3},o),(i_{4},o),(i_{6},OBC),(i_{7},ST),(i_{8},SC)\}.
 $$
 
Note that $\eta^{FT}(I,T)\setminus \eta^{BT}(I,T)=\{(i_{5},o)\}$ and $\eta^{BT}(I,T)\setminus \eta^{FT}(I,T)=\{(i_{8},SC)\}$. 
	
\end{example}

\subsection{Merit-Optimality and DA Compatibility}

In Example \ref{ex:BTvFT}, the set of individuals selected by $\eta^{FT}(I,T)$ is a more meritorious set than the set of individuals chosen induced by $\eta^{BT}(I,T)$. We now generalize this observation and establish that the FT choice rule merit-dominates other choice rules induced by the assignment rules in the family $\mathbb{A}$. 

\begin{theorem}\label{thm:FT}
$\eta^{FT}$ merit-dominates any other assignment rule in $\mathbb{A}$. 
\end{theorem}

The proof of Theorem~4 is not a straightforward application of existing results on the precedence order of reserve categories. Different transfer schemes may generate different cumulative numbers of de-reserved positions and, hence, different terminal capacity vectors. Thus, a comparison of precedence orders while holding capacities fixed is insufficient to establish the result. The key
step in our proof is to establish that every applicant selected under the initial hard-reserve allocation remains selected under every admissible transfer
scheme, although the types of positions assigned to these applicants may change. This containment property, together with an upper bound on the number of additional selections, enables us to establish merit-domination.

This theorem is the paper's main implementation result for \textit{P.V. Indiresan} (2011). Once the Court fixes OBC eligibility exogenously and permits conversion only after eligible OBC candidates are exhausted, forward transfer is the most meritorious way to implement the resulting de-reservation. Open positions select a weakly more meritorious residual cohort when they are processed later, because high-scoring protected-category applicants remain available to satisfy their category-specific reserves before the converted open positions are filled.

\begin{theorem}\label{thm:family}
    The choice rule induced by every assignment rule in the family $\mathbb{A}$ is consistent and satisfies the substitutes property and size monotonicity. 
\end{theorem}

The proof of Theorem~5 presents a distinct technical challenge. When the applicant pool expands, not only does the competition for existing positions change, but the cumulative number of de-reserved positions and their distribution across the precedence order may also change. Standard arguments establishing
substitutability and size monotonicity for sequential choice rules with fixed capacities therefore do not directly apply. To overcome this difficulty, we
characterize the terminal number of transferred positions as the smallest solution to a vacancy-free condition and establish monotonicity properties of
the residual applicant pools under simultaneous changes in applicant populations and position capacities. These intermediate results are instrumental in establishing consistency, substitutability, and size monotonicity.

Theorem \ref{thm:family} establishes the admissibility of any assignment rule in the family $\mathbb{A}$ for institutions in centralized markets.\footnote{The results established by \cite{hatfield2005matching} ensure that the deferred acceptance mechanism in conjunction with any assignment rule in the family $\mathbb{A}$ is stable and strategy-proof.}

\section{Comparing the JNU and P.V. Indiresan Interpretations}\label{sec:comparelegal}

We finally compare the two legal interpretations while holding fixed the de-reservation principle that vacant OBC positions are transferred forward. The rule $\eta^{JNU}$ uses the JNU interpretation. After the original open positions are filled, an OBC applicant is eligible only if her score weakly exceeds the endogenous threshold $\theta^{o}(A)-\lambda$. The rule $\eta^{FT}$ instead uses the predetermined \textit{P.V. Indiresan} threshold $\underline{\theta}^{OBC}$. Both rules process de-reserved positions last. Their only substantive difference is therefore the set of OBC applicants who are eligible before conversion.

Define the JNU threshold by $\tau^{JNU}(A)=\theta^{o}(A)-\lambda$ and the \textit{P.V. Indiresan} threshold by $\tau^{PVI}=\underline{\theta}^{OBC}$. If $\tau^{JNU}(A)\geq \tau^{PVI}$, the JNU interpretation is weakly more restrictive in the sense that every OBC applicant eligible under $\eta^{JNU}$ is also eligible under $\eta^{FT}$, but the converse need not hold. Consequently, $\eta^{JNU}$ may assign fewer positions as OBC positions and transfer more positions to merit-based open competition. When the two rules choose different cohorts, this can only replace an OBC applicant admitted under the weaker \textit{P.V. Indiresan} threshold with a weakly higher-scoring remaining applicant selected through the final open transfer.

\begin{proposition}\label{prop:JNU_FT_cardinality}
    For every set of applicants $A\subseteq I$ and every membership profile $T\in\mathcal{T}$,
\[
|C^{JNU}(A)|=|C^{FT}(A)|.
\]
\end{proposition}

This equality holds irrespective of the relative stringency of the endogenous and exogenous OBC eligibility thresholds.

Proposition~\ref{prop:JNU_FT_cardinality} establishes that the two legal interpretations, when implemented through forward transfers, are equivalent in terms of \textbf{aggregate position utilization}. Thus, neither interpretation
has an advantage over the other in terms of the total number of applicants admitted. Their differences concern the composition of the selected cohort, the distribution of applicants across position types, and the structural
properties of the induced choice rules. In particular, the merit comparison in the following theorem is a comparison of cohorts of identical cardinality.

\begin{theorem}\label{thm:etaJNUFT}
Fix an applicant set $A$. Suppose that $\tau^{JNU}(A)\geq \tau^{PVI}$. Then the two selected cohorts coincide or the cohort selected by $\eta^{JNU}$ merit-dominates the cohort selected by $\eta^{FT}$. 
\end{theorem}

The condition in Theorem~\ref{thm:etaJNUFT} has a natural institutional interpretation. It requires that the score of the last applicant admitted to an original open-category position exceed the predetermined OBC qualifying threshold
by at least $\lambda$. This condition is particularly plausible in highly selective admissions environments, where the realized admission cutoffs substantially exceed minimum qualifying scores. For example, the official JEE Advanced 2025 report specifies minimum aggregate qualifying marks of 74 for the common rank list and 66 for the OBC-NCL rank list, out of 360 possible marks. In contrast, admission to many IIT programs requires considerably higher examination scores. These observations suggest that the condition in Theorem~\ref{thm:etaJNUFT} is empirically relevant for competitive higher-education admissions. Under this condition, the JNU interpretation weakly increases the merit of the selected cohort relative to the exogenous cutoff established in \textit{P.V. Indiresan} (2011).

Note that if $\tau^{JNU}(A)<\tau^{PVI}$ for some applicant pool, no unconditional merit-dominance ranking between $\eta^{JNU}$ and $\eta^{FT}$ follows from the legal restrictions alone.

The theorem separates two conceptually different questions. Within either legal interpretation, forward transfer is the merit-optimal treatment of OBC vacancies. However, in all interpretations, merit depends on which eligibility threshold is more demanding. The 2011 judgment therefore changes not the optimal \textit{location} of de-reserved positions in the precedence order, but the set of OBC applicants who must be considered before de-reservation is legally permissible.

\section{Related Literature}\label{sec:rellit}

Our paper contributes to the growing theoretical literature on affirmative action in matching markets. \cite{abdulkadiroglu2003school} introduced the majority quota in the context of school choice  to promote diversity. \cite{kojima2012school} showed that strengthening the majority quotas can paradoxically harm all members of the intended beneficiaries in any stable mechanism. \cite{hafalir2013effective} demonstrated that replacing the majority quotas with the soft minority reserves resolves this perversity. \cite{ehlers2014school} extended this analysis to systems with multiple protected categories. The idea of de-reservation in the form of capacity transfers between position types was first formally introduced in \cite{westkamp2013analysis}. \cite{kominers2016matching} introduced the slot-specific priorities approach in a matching model with contracts. \cite{aygun2020dynamic} generalize the idea of capacity transfer in this framework. \cite{avataneo2021slot} introduce a family of choice rules based on slot-specific priorities with capacity transfers.

Beyond reserve size, \cite{dur2018reserve,dur2020explicit} establish that the precedence order in which different seat types are filled constitutes an independent policy instrument with substantial distributional consequences. \cite{almeer2024increasing} further investigate how precedence orders can be modified to increase the representation of targeted groups in general reserve systems. The practical importance of reserve sequencing is reinforced by \cite{pathak2023reversing}, who document systematic misunderstandings of its distributional implications. Similarly, \cite{pathak2024fair} develop a general theory of reserve design in the context of healthcare rationing, emphasizing the role of eligibility restrictions and sequential reserve processing in reconciling competing allocation objectives.

Our merit-optimality results build on this precedence-order logic but address a fundamentally different theoretical problem. In our framework, the capacities of different position types are determined endogenously through an iterative transfer procedure. Consequently, alternative transfer schemes may induce different terminal capacity vectors, as well as different allocation sequences. Comparisons of precedence orders under fixed capacities are therefore insufficient to establish our results. Our analysis develops a different proof methodology based on structural properties of sequential selection under endogenous capacity adjustments and identifies the transfer scheme that merit-dominates every alternative admissible implementation.

\cite{echenique2015control} is the first paper to present Indian affirmative action as an example of controlled school choice. \cite{aygun2017large} explore the specific challenges faced in admissions to the Indian Institutes of Technology (IIT).  \cite{aygun2020dynamic} and \cite{aygun2026matching} offer a comprehensive treatment of vertical reservations and de-reservations by introducing a family of \textit{dynamic reserves} and \textit{generalized sequential} choice rules, respectively. \cite{sonmez2024constitutional} study the interaction of horizontal and vertical reservations in centralized matching markets with multiple institutions. \cite{aygun2026IndiaMD} study the interaction with horizontal and vertical reservations when horizontal reservations adhere to the hierarchical structure by adopting a one-to-all matching policy.

\cite{evren2026affirmative} examine the implementation of the Central Educational Institutions (Reservation in Teachers' Cadre) Act, 2019, which regulates reservations in faculty recruitment. Their central problem is how to implement fractional reservation entitlements when indivisible positions arrive sequentially across university departments, and appointments are irreversible. \cite{evren2026affirmative} study how reserved capacities should be determined and implemented over time, while our paper takes initial reserved capacities as given and studies how unfilled reserved capacity should be reallocated among position types.

Beyond India, several countries, including Brazil and Chile, have implemented large-scale affirmative action and diversity policies in school and college admissions. \cite{aygun2021college} study multidimensional reserves in Brazilian university admissions and identify incentive and fairness problems arising from applicants' strategic choices of affirmative-action privileges. In Chile, \cite{correa2019school} examine the design and implementation of centralized school choice, while \cite{rios2021flexible} study the integration of affirmative-action policies into centralized college admissions and identify inefficiencies associated with flexible quotas. More recently, \cite{escobar2026minority} investigated the effects of minority reserves on school segregation and allocative efficiency using a theoretical model and data from Chile. Complementing these market-design contributions, \cite{barahona2021affirmative} study the distributional consequences of affirmative action in centralized Brazilian college admissions, while \cite{mello2023affirmative} examines how higher-education quota influence students' earlier school choices. These studies illustrate the importance of allocation mechanisms in determining the effectiveness and distributional consequences of affirmative-action policies. Our paper contributes to this literature by providing an axiomatic analysis of eligibility thresholds and de-reservation policies.

Our paper is also related to \citet{evren2026reserve}, who develops a general framework for reserve systems with eligibility thresholds, allowing independent category-specific priorities, overlapping categories, and mixtures of hard and soft reserves. The author proposes the Iterative Max-in-Max Assignment Mechanism, which maximizes resource utilization and subsequently lexicographically maximizes beneficiary assignments while satisfying path independence. Our paper allows the capacities of different position types to adjust endogenously through iterative transfers and establishes that forward transfers merit-dominate alternative de-reservation schemes. Thus, while \cite{evren2026reserve} focuses on utilization and beneficiary accommodation under general threshold reserves, our analysis identifies merit-optimal de-reservation policies subject to legally mandated eligibility and allocation constraints.

Another related paper is \cite{yilmaz2023equity}, which considers the allocation of scarce and identical objects through reserve categories, motivated in part by the equitable rationing of medical resources such as vaccines and ventilators. His central objective is to characterize random allocation rules that mitigate unequal treatment of agents. Instead, our paper studies deterministic merit-based assignment rules subject to the legal requirements governing affirmative action in India.

Our analysis also contributes to the axiomatic literature on choice rules with affirmative action. \cite{dougan2022choice} characterize monotonic and substitutable affirmative action choice rules through bounded reserve representations, while \cite{dur2023fairness} investigate fairness and score-wise undominatedness in reserve systems with overlapping beneficiary categories. Complementing these contributions, we derive axioms from judicial mandates and examine how alternative interpretations of eligibility requirements and de-reservation policies affect merit-optimality, consistency, substitutability, and size monotonicity. In particular, our results demonstrate that the interaction between endogenous eligibility thresholds and the reallocation of unused reserved capacity can generate complementarities even when each component, considered separately in our environment, is compatible with substitutability.

Like ours, several articles in the literature, such as \cite{abdulkadiroglu2025market}, \cite{sonmez2022affirmative}, \cite{arnosti2026explainable}, \cite{aygun2026IndiaMD}, among many others, use Gale-domination criteria to compare different allocation rules under affirmative action.

\section{Conclusion}\label{sec:concl}

In this paper, we develop an axiomatic analysis of India's vertical reservation system following the evolution of the Supreme Court's jurisprudence from Indra Sawhney (1992), through Ashoka Kumar Thakur (2008), to P.V. Indiresan (2011). We formalize the principles established by these judgments and characterize the assignment procedures associated with their different interpretations of eligibility and de-reservation. Our analysis identifies the difficulties generated by endogenous eligibility cutoffs, including the potential incompatibility of legal requirements and the failure of substitutability. These findings provide a theoretical justification for the interpretation adopted in P.V. Indiresan (2011), under which eligibility thresholds are determined in advance rather than by the realized allocation.

Our main contribution concerns the implementation of de-reservation under this legally operative framework. When eligible OBC applicants are not sufficient to fill reserved positions, the remaining positions must be converted to open-category positions. Although this requirement determines when conversion is permissible, it does not uniquely specify where the converted positions should be placed in the allocation sequence. We introduce a family of assignment rules that differ in their treatment of these positions and establish that the forward-transfer rule, which processes all converted positions after the original reserved categories, merit-dominates every other rule in this family. Thus, even when reservation quotas, eligibility requirements, and the total number of positions are held fixed, the sequencing of de-reservation can improve the merit of the selected cohort. This result yields a concrete policy recommendation. 

More broadly, our findings illustrate how market design can complement legal and institutional frameworks. Judicial decisions establish the objectives and constraints that govern affirmative action, but they need not resolve every operational detail of an allocation procedure. Such details are not merely administrative in the sense that different implementations of the same legal requirements can produce systematically different outcomes. By identifying the merit-optimal procedure within a family of admissible rules, our analysis shows how economic theory can translate legal principles into precise and implementable allocation mechanisms. The broader lesson extends beyond India's reservation system. When affirmative action policies allow for the reallocation of unused reserved capacity, the sequencing of that reallocation is itself an important instrument of market design.

\newpage
\bibliographystyle{aer}
\bibliography{IndianAA}

\newpage

\section{ Online Appendix}

\section*{Proof of Theorem \ref{thm:IS}}
($\Longleftarrow$) By the definition of $\eta^{IS}$, the first $q^{o}$ highest-scoring applicants are assigned open-category positions. If there are fewer than $q^{o}$ applicants, all are assigned open-category positions. Thus, the OA principle is satisfied. 

Since each category fills its positions following the merit score rankings, the WCF property is trivially satisfied. The only way for an individual $i$ to be unable to be assigned is that the positions $i$ is eligible for are filled with individuals with higher merit scores. Thus, the QFE principle is also satisfied.

($\Longrightarrow$) Let $\eta$ be an assignment rule that satisfies the three axioms. By the OA principle, we have $\{(i,o):(i,o)\in \eta (A,T)\}=\{(i,o):(i,o)\in \eta^{IS}\left(A,T\right)\}$.
If $\mid A\mid\leq q^{o}$, then we are done. Let $\mid A\mid>q^{o}$. The QFE requires that if an individual $i$ with $t\left(i\right)\in R$ is rejected, then $\mid C^{t\left(i\right)}\mid=q^{t\left(i\right)}$.
The QFE and WCF jointly imply that $C^{v}$ chooses the top $q^{v}$ individuals of the set $\left(A\setminus C^{o}\left(A\right)\right)\bigcap A^{v}$,
for all $v\in R$. Thus, $\eta(A,T)=\eta^{IS}(A,T)$.

\section*{Proof of Theorem \ref{thm:AKT}}
\emph{($\Leftarrow$)}  
By construction, $\eta^{AKT}$ satisfies the OA principle: in Stage 1 it assigns the
$q^{o}$ highest-scoring applicants in $A$ to the open-category positions, or all applicants if $|A|<q^{o}$.

$\eta^{AKT}$ also satisfies the WCF. Indeed, in Stage 1 the open-category positions are assigned in decreasing order of score. In Stage 2, for each $r\in R\setminus\{OBC\}$, the category-$r$ positions are assigned based on merit scores among the remaining applicants of type $r$. In Stage 3, the OBC slots are assigned based on merit scores among the remaining OBC applicants whose scores satisfy the cutoff requirement. Hence, no lower-scoring applicant of a given type is assigned unless every higher-scoring applicant of the same type is also assigned.

Finally, $\eta^{AKT}$ satisfies mQFE by the definition of Stages 2 and 3. For each
$r\in R\setminus\{OBC\}$, all category-$r$ positions are filled whenever there are enough remaining applicants of type $r$. Otherwise, all such applicants are selected. For the OBC category, Stage 3 assigns the highest-scoring remaining OBC applicants whose scores are at least the open-category cutoff minus $\lambda$, up to capacity $q^{OBC}$; thus, if an eligible OBC applicant remains unmatched, then all OBC positions must already be filled, and no ineligible OBC applicant is assigned.

\emph{($\Rightarrow$)}  
Let $\eta$ be any assignment rule satisfying the OA principle, WCF, and mQFE. We show
that $\eta=\eta^{AKT}$. Fix a set of applicants $A$ and a category profile $T=(t(i))_{i\in A}$. We will prove that $\eta(A,T)=\eta^{AKT}(A,T)$.

By the OA principle, $\{\,i\in A:(i,o)\in \eta(A,T)\,\}= \{\,i\in A:\operatorname{rank}_{A}(i)\le q^{o}\,\}$. This is exactly the set assigned to the open-category in Stage 1 of $\eta^{AKT}$. Hence, the open-category assignments under $\eta$ and $\eta^{AKT}$ coincide.

Let $A^{1}=A\setminus \{\,i\in A:(i,o)\in \eta(A,T)\,\}$. Since the open-category assignments under $\eta$ and $\eta^{AKT}$ are identical, the set $A^{1}$ is also exactly the set of applicants remaining after Stage 1 of $\eta^{AKT}$. Fix $r\in R\setminus\{OBC\}$, and let $A^{1}_r=\{\,i\in A^1: t(i)=r\,\}$ be the set of remaining applicants of type $r$ after seats in the open-category are filled.

We claim that under $\eta$, the applicants assigned to category $r$ are exactly the highest-scoring $\min\{|A^1_r|,q^{r}\}$ applicants in $A^1_r$. 

First, suppose that $|A^1_r|\le q^{r}$. Then every applicant in $A^1_r$ must be assigned to category-$r$. Otherwise, if some $i\in A^1_r$ were not assigned, then since $t(i)=r\neq OBC$, part (1) of mQFE would require that all $q^{r}$ category-$r$ seats be filled. But there are fewer than $q^{r}$ applicants of type $r$ remaining, which is impossible because category-$r$ positions can only be assigned to applicants of type $r$. Hence, all applicants in $A^1_r$ must be assigned.

Now suppose that $|A^1_r|>q^{r}$. Then exactly $q^{r}$ applicants of type $r$ must be
assigned, again by part (1) of mQFE. Moreover, WCF implies that these $q^{r}$ assignees must be the highest-scoring $q^{r}$ applicants in $A^1_r$: if a lower-scoring applicant $j\in A^1_r$ were assigned while a higher-scoring applicant $i\in A^1_r$ were not assigned, then WCF would be violated. Thus, for every $r\in R\setminus\{OBC\}$, the category-$r$ assignments under $\eta$ coincide with those in Stage 2 of $\eta^{AKT}$.

Let $A^1_{OBC}=\{\,i\in A^1: t(i)=OBC\,\}$ be the set of OBC applicants remaining after the open-category seats are filled. Let $\underline\theta^o(A)=\min\{\theta(j):(j,o)\in \eta(A,T)\}$ be the open-category cutoff score. Since open-category assignments are already determined by OA, this is exactly the same cutoff score used in Stage 3 of $\eta^{AKT}$.

Let $E=\{\,i\in A^1_{OBC}: \theta(i)\ge \underline\theta^o(A)-\lambda\,\}$ be the set of remaining eligible OBC applicants. We now show that under $\eta$, the applicants assigned to the OBC category are exactly the highest-scoring $\min\{|E|,q^{OBC}\}$ applicants in $E$.

Take any $i\in A^1_{OBC}\setminus E$. Then $\theta(i)<\underline\theta^o(A)-\lambda$. By part (3) of mQFE, such an applicant cannot be assigned to an OBC slot. Hence
every OBC assignee under $\eta$ must belong to $E$. Thus, no ineligible OBC applicant can be assigned.

If $|E|\le q^{OBC}$, then every applicant in $E$ must be assigned. Suppose, to the contrary, that some $i\in E$ is not assigned. Since $i$ is eligible and unassigned, part (2) of mQFE implies that all $q^{OBC}$ OBC seats must be filled. But by the preceding paragraph, only applicants in $E$ can be assigned to OBC seats, and $|E|\le q^{OBC}$. Hence, it is impossible to fill all $q^{OBC}$ OBC seats while leaving some applicants in $E$ unmatched. Contradiction. Therefore, every applicant in $E$ must be assigned.

If $|E|>q^{OBC}$, then exactly the highest-scoring $q^{OBC}$ applicants in $E$
must be assigned. Because there are more than $q^{OBC}$ eligible OBC applicants, capacity implies that at most $q^{OBC}$ of them can be assigned. On the other hand, if some eligible OBC applicants were left unmatched, part (2) of mQFE requires that all $q^{OBC}$ OBC seats be filled. Thus, exactly $q^{OBC}$ applicants must be assigned to OBC seats.

Finally, WCF implies that these assignees must be the highest-scoring $q^{OBC}$
applicants in $E$: if a lower-scoring eligible OBC applicant $j\in E$ were assigned while a higher-scoring eligible OBC applicant $i\in E$ were not assigned, then $t(i)=t(j)=OBC$ and $i\succ j$, contradicting WCF.

We can conclude that the OBC assignments under $\eta$ coincide exactly with the OBC
assignments in Stage 3 of $\eta^{AKT}$. Therefore, $\eta(A,T)=\eta^{AKT}(A,T)$. Since $A$ and $T$ were arbitrary, it follows that $\eta=\eta^{AKT}$.

\section*{Proof of Proposition \ref{prop:JNU}}

We first show that $C^{AKT}$ satisfies the substitutes condition. Consider
$i, j \in I$ and $A \subseteq I \setminus \{i,j\}$ such that
$i \notin C^{AKT}(A \cup \{i\})$. We need to show that
$i \notin C^{AKT}(A \cup \{i,j\})$.

Let $B = A \cup \{i\}$ and $B' = A \cup \{i,j\}$. For $X \in \{B, B'\}$, let
$C^o(X) = \{k \in X : rank_X(k) \leq q^o\}$ be the set of applicants assigned
open-category positions in Stage 1 of $\eta^{AKT}(X,T)$, let
$X^1 = X \setminus C^o(X)$ be the set of applicants remaining after Stage 1,
and let $\theta^o(X) = \min\{\theta(k) : k \in C^o(X)\}$ be the open-category
cutoff.

Since $i \notin C^{AKT}(B)$, applicant $i$ is not assigned an open-category
position, so $i \in B^1$ and $|B| > q^o$. We use the following two facts.

Note that $B^1 \subseteq (B')^1$. If $k \in B^1$, then
at least $q^o$ applicants in $B$ have a higher score than $k$. All of them
belong to $B'$, so $k \notin C^o(B')$.

Moreover, note that $\theta^o(B') \geq \theta^o(B)$. Since
$|B'| > |B| > q^o$, both $C^o(B)$ and $C^o(B')$ contain exactly $q^o$
applicants. $B^1 \subseteq (B')^1$ implies $C^o(B') \subseteq C^o(B) \cup \{j\}$. Hence, $C^o(B') = C^o(B)$ or $C^o(B')$ is obtained from $C^o(B)$ by replacing its lowest-scoring member with $j$, in which case $j$ has a higher score than the replaced applicant. In both cases, the minimum score weakly increases.

Since $B^1 \subseteq (B')^1$, $i \in (B')^1$, so $i$ is not assigned an open-category position in $\eta^{AKT}(B',T)$. We consider three cases.

\smallskip
\noindent\textit{Case 1: $t(i) = g$.} Applicant $i$ is eligible only for
open-category positions. Since $i \in (B')^1$, we have
$i \notin C^{AKT}(B')$.

\smallskip
\noindent\textit{Case 2: $t(i) = r \in \mathcal{R} \setminus \{OBC\}$.} Since
$i \in B^1$ is not selected in Stage 2 of $\eta^{AKT}(B,T)$, there are at least
$q^r$ applicants of category $r$ in $B^1$ with a higher score than $i$. Since $B^1 \subseteq (B')^1$, all of them belong to $(B')^1$. Hence at least $q^r$ applicants of
category $r$ in $(B')^1$ have a higher score than $i$, and $i$ is not selected
in Stage 2 of $\eta^{AKT}(B',T)$. Therefore, $i \notin C^{AKT}(B')$.

\smallskip
\noindent\textit{Case 3: $t(i) = OBC$.} If
$\theta(i) < \theta^o(B') - \lambda$, then $i$ is ineligible for an OBC
position in $\eta^{AKT}(B',T)$, and hence $i \notin C^{AKT}(B')$.

Suppose instead that $\theta(i) \geq \theta^o(B') - \lambda$. Since $\theta^o(B') \geq \theta^o(B)$, we have $\theta(i) \geq \theta^o(B) - \lambda$, so $i$ is eligible for an OBC position
in $\eta^{AKT}(B,T)$. Since $i \in B^1$ is eligible but not selected in Stage 3
of $\eta^{AKT}(B,T)$, there are at least $q^{OBC}$ OBC applicants in $B^1$ with
a higher score than $i$. Let $k$ be any such applicant. Since $B^1 \subseteq (B')^1$,
$k \in (B')^1$. Moreover,
\[
\theta(k) > \theta(i) \geq \theta^o(B') - \lambda ,
\]
so $k$ is eligible for an OBC position in $\eta^{AKT}(B',T)$. Thus at least
$q^{OBC}$ eligible OBC applicants in $(B')^1$ have a higher score than $i$, and
$i$ is not selected in Stage 3 of $\eta^{AKT}(B',T)$. Therefore,
$i \notin C^{AKT}(B')$.

\smallskip
In all cases, $i \notin C^{AKT}(A \cup \{i,j\})$. Hence $C^{AKT}$ satisfies the
substitutes condition.

We next show that $C^{AKT}$ satisfies consistency.
Fix $A\subseteq I$ and $i\in I\setminus A$, and let
$B=A\cup\{i\}$. Suppose that $i\notin C^{AKT}(B)$. Since $i$ is not selected in Stage 1 of $\eta^{AKT}(B,T)$, removing $i$ does not affect the set of applicants assigned open-category positions. Thus, $C^o(A)=C^o(B)$. In particular, the open-category cutoff remains unchanged. That is, $\theta^o(A)=\theta^o(B)$.

For each $r\in R\setminus\{OBC\}$, the removal of $i$
does not affect the category-$r$ selection. Indeed,
if $t(i)\neq r$, the set of applicants considered
for category-$r$ positions is unchanged. If $t(i)=r$,
then $i$ is not among the applicants selected in
category $r$ under $\eta^{AKT}(B,T)$, so removing
$i$ does not affect the merit-based selection in
that category.

Finally, consider the OBC category. Since the
open-category cutoff is unchanged, the OBC eligibility
threshold is also unchanged. If $i$ is not an eligible
OBC applicant, removing $i$ does not affect the set of
eligible OBC applicants. If $i$ is an eligible OBC
applicant, then $i\notin C^{AKT}(B)$ implies that
all OBC positions are assigned to eligible applicants
with higher priority than $i$. Removing $i$ therefore
does not affect the OBC selection.

Thus, the selections in all three stages
remain unchanged, and hence
\[
C^{AKT}(A)=C^{AKT}(B).
\]
Since $A$ and $i$ were arbitrary, $C^{AKT}$
satisfies consistency.

\section*{Proof of Theorem \ref{thm:etaJNU}}

We first establish that $\eta^{JNU}$ satisfies the four axioms. Fix an arbitrary set of applicants $A\subseteq I$ and a category membership profile $T\in\mathcal{T}$.

By construction, $\eta^{JNU}$ assigns the highest-scoring $\min\{|A|,q^o\}$ applicants to the original open-category positions in Stage 1. Thus, $\eta^{JNU}$ satisfies the OA$^{*}$ principle.

To establish WCF$^{*}$, consider any two applicants $i,j\in A$ such that $i\succ j$ and $(j,\tau)\in\eta^{JNU}(A,T)$ for some $\tau\in\{o,d\}\cup R$. If $\tau=o$, then $i$ must also have been assigned an original open-category position in Stage 1.

If $\tau=d$, then either $i$ was assigned a position
in one of the first three stages or $i$ remained
available at the beginning of Stage 4. In the latter
case, since category-$d$ positions are assigned in
decreasing order of merit and $i\succ j$, applicant
$i$ must also have been assigned a category-$d$
position.

Finally, suppose that $\tau\in R$ and $t(i)=t(j)$.
If $i$ was not assigned an original open-category
position, then $i$ was available when the positions
reserved for category $\tau$ were processed.
If $\tau\neq OBC$, the merit-based selection within
category $\tau$ implies that $i$ must have been
assigned a position in that category.
If $\tau=OBC$, the eligibility of $j$ implies
\[
\theta(i)\geq\theta(j)
\geq\theta^o(A)-\lambda,
\]
where $\theta^o(A)$ denotes the score of the
lowest-scoring applicant assigned an original
open-category position. Thus, $i$ is also eligible
for an OBC position. Since $i\succ j$, applicant $i$
must have been assigned an OBC position.

In all cases, $i$ is assigned a position.
Therefore, $\eta^{JNU}$ satisfies WCF$^{*}$.

Next, consider mQFE. For every reserve category
$r\in R\setminus\{OBC\}$, Stage 2 fills all
category-$r$ positions whenever sufficiently many
applicants of category $r$ remain after Stage 1.
Otherwise, all remaining applicants of category $r$
are assigned positions. Similarly, Stage 3 fills
all OBC positions whenever sufficiently many
eligible OBC applicants remain after Stage 1.
Otherwise, all remaining eligible OBC applicants
are assigned positions. Moreover, no OBC applicant
whose score falls below $\theta^o(A)-\lambda$
is assigned an OBC position. The subsequent
assignment of category-$d$ positions does not
alter any of these assignments or the original
open-category cutoff. Hence, $\eta^{JNU}$
satisfies mQFE.

Finally, let $\kappa$ denote the number of vacant
OBC positions at the end of Stage 3. Stage 4
assigns up to $\kappa$ category-$d$ positions
to the highest-scoring remaining applicants.
If any of these positions remain vacant, then
every applicant in $A$ has been assigned a
position. Therefore, $\eta^{JNU}$ satisfies
OBC de-reservation$^{*}$.

Notice also that, by construction,
\[
\left|\left\{(i,\tau)\in\eta^{JNU}(A,T):
\tau\in\{d,OBC\}\right\}\right|
\leq q^{OBC}.
\]
Thus, $\eta^{JNU}$ satisfies the combined capacity
constraint in Definition~9.

We next establish the merit-optimality of $\eta^{JNU}$ among all assignment rules satisfying these axioms. Let $\eta'$ be an arbitrary assignment rule satisfying
the OA$^{*}$ principle, WCF$^{*}$, mQFE, and OBC de-reservation$^{*}$. Let $C'$ denote its induced choice rule.

We need to show that  $C^{JNU}(A)\succeq_M C'(A)$ for every $A\subseteq I$ and $T\in\mathcal{T}$. If $|A|\leq q^o$, the OA$^{*}$ principle requires
both assignment rules to assign every applicant an original open-category position. Hence, $C^{JNU}(A)=C'(A)=A$. We, thus, restrict our attention to the case $|A|>q^o$.

\begin{claim}
    The original open-category assignments under $\eta^{JNU}$ and $\eta'$ are identical.
\end{claim}

\begin{proof}[Proof of Claim 1]
  Let $C^o(A)= \left\{i\in A:\operatorname{rank}_A(i)\leq q^o\right\}$ denote the set of the $q^o$ highest-scoring applicants in $A$. By construction, $\eta^{JNU}$ assigns precisely the members of $C^o(A)$ to original open-category positions.

We show that $\eta'$ does the same. By the
OA$^{*}$ principle, exactly $q^o$ applicants
are assigned original open-category positions
under $\eta'$.

Suppose, toward a contradiction, that some
$i\in C^o(A)$ is not assigned an original
open-category position under $\eta'$. Since
exactly $q^o$ applicants receive original
open-category positions, there exists an
applicant $j\notin C^o(A)$ such that
\[
(j,o)\in\eta'(A,T).
\]
By the definition of $C^o(A)$, we have $i\succ j$.

Since $j$ is assigned an original open-category
position, WCF$^{*}$ requires that $i$ also be
assigned a position. By hypothesis, this position
is not an original open-category position.
Consequently, there exists some
$\tau\in\{d\}\cup R$ such that
\[
(i,\tau)\in\eta'(A,T).
\]
However, the OA$^{*}$ principle requires every
applicant assigned an original open-category
position to have higher priority than every
applicant assigned a reserved or category-$d$
position. Therefore, $j\succ i$, contradicting
$i\succ j$.

It follows that
\[
\{i\in A:(i,o)\in\eta'(A,T)\}=C^o(A).
\]
Hence, the original open-category assignments under $\eta^{JNU}$ and $\eta'$ are identical.  
\end{proof}

Thus, the two assignment rules induce
the same original open-category cutoff, $\theta^o(A)=\min_{i\in C^o(A)}\theta(i)$,
and hence the same OBC eligibility threshold, $\theta^o(A)-\lambda$.

Let $A^1=A\setminus C^o(A)$ denote the set of applicants who are not assigned
original open-category positions.

\begin{claim}
Every applicant selected in Stage 2 of $\eta^{JNU}$ is selected under $\eta'$.
\end{claim}

\begin{proof}[Proof of Claim 2]
Fix an arbitrary reserve category $r\in R\setminus\{OBC\}$ and define
\[
A^{1,r}=\{i\in A^1:t(i)=r\}.
\]
Let $C^r(A^1)$ denote the set of the
$\min\{q^r,|A^{1,r}|\}$ highest-scoring
applicants in $A^{1,r}$.

By construction, $C^r(A^1)$ is the set of
applicants assigned category-$r$ positions
in Stage 2 of $\eta^{JNU}$.

We claim that $C^r(A^1)\subseteq C'(A)$. Suppose, toward a contradiction, that there exists an applicant
\[
i\in C^r(A^1)\setminus C'(A).
\]
Since $i$ belongs to category $r$ and remains
unassigned under $\eta'$, part (1) of mQFE
requires that all $q^r$ category-$r$ positions
be assigned. That is,
\[
\left|\left\{j\in A:(j,r)\in\eta'(A,T)\right\}\right|
=q^r.
\]

By Claim 1 and the feasibility conditions of assignment rules, every applicant assigned a category-$r$ position under $\eta'$ belongs
to $A^{1,r}$.

Since $i\in C^r(A^1)$, there are at most
$q^r-1$ applicants in $A^{1,r}$ with higher
priority than $i$. Therefore, among the $q^r$
applicants assigned category-$r$ positions
under $\eta'$, there must exist an applicant
$j$ such that
\[
t(j)=t(i)=r,\qquad i\succ j,
\qquad (j,r)\in\eta'(A,T).
\]

By WCF$^{*}$, applicant $i$ must also be assigned a position under $\eta'$. This
contradicts $i\notin C'(A)$. Thus, $C^r(A^1)\subseteq C'(A)$ for every $r\in R\setminus\{OBC\}$.
\end{proof}

Note that this inclusion does not require the applicants in $C^r(A^1)$ to receive category-$r$ positions under $\eta'$. Some of them may instead receive
category-$d$ positions.

\begin{claim}
    Every applicant selected in Stage 3 of $\eta^{JNU}$ is selected under $\eta'$.
\end{claim}

\begin{proof}[Proof of Claim 3]
Define the set of remaining eligible OBC applicants by
\[
E^{OBC}(A)=
\left\{
i\in A^1:
t(i)=OBC,\ 
\theta(i)\geq\theta^o(A)-\lambda
\right\}.
\]

Let $m=\min\{q^{OBC},|E^{OBC}(A)|\}$, and let $C^{OBC}(A^1)$ denote the set of
the $m$ highest-scoring applicants in $E^{OBC}(A)$. By construction, $C^{OBC}(A^1)$ is the set of applicants assigned OBC positions in Stage 3 of $\eta^{JNU}$.

We claim $C^{OBC}(A^1)\subseteq C'(A)$. Suppose, toward a contradiction, that there exists an applicant $i\in C^{OBC}(A^1)\setminus C'(A)$.

Since $i\in E^{OBC}(A)$ and the original
open-category cutoff is identical under
$\eta^{JNU}$ and $\eta'$, applicant $i$
is eligible for an OBC position under
$\eta'$.

Because $i$ remains unassigned, part (2)
of mQFE requires that
\[
\left|\left\{j\in A:
(j,OBC)\in\eta'(A,T)\right\}\right|
=q^{OBC}.
\]

By part (3) of mQFE, every applicant assigned an OBC position under $\eta'$ satisfies the OBC eligibility threshold. Moreover, by Claim 1, none of these
applicants belongs to $C^o(A)$. Hence, every applicant assigned an OBC
position under $\eta'$ belongs to $E^{OBC}(A)$.

Since $i\in C^{OBC}(A^1)$, there are at
most $q^{OBC}-1$ applicants in
$E^{OBC}(A)$ with higher priority than
$i$. Therefore, there exists an applicant
$j$ such that
\[
t(i)=t(j)=OBC,\qquad i\succ j,
\qquad (j,OBC)\in\eta'(A,T).
\]

By WCF$^{*}$, applicant $i$ must also be assigned a position under $\eta'$, contradicting $i\notin C'(A)$. Thus, $C^{OBC}(A^1)\subseteq C'(A)$. 

\end{proof}

\begin{claim}
  $B_0\subseteq C'(A)$.  
\end{claim}

\begin{proof}[Proof of Claim 4]

Let $B_0= C^o(A) \cup \bigcup_{r\in R\setminus\{OBC\}}C^r(A^1) \cup C^{OBC}(A^1)$.

The set $B_0$ consists of all applicants selected in the first three stages of
$\eta^{JNU}$. Equivalently, $B_0=C^{AKT}(A)$. Combining Claims 1--3, we obtain $B_0\subseteq C'(A)$.
\end{proof}

Claim 4 established that every applicant selected by $\eta^{AKT}$ must be selected under any assignment rule satisfying the four axioms, although the types of positions assigned to these applicants may differ.

\begin{claim}
    Any assignment rule satisfying the four axioms can select at most $\kappa$ applicants in addition to the members of $B_0$, where $\kappa=q^{OBC}-m$ denote the number of vacant OBC positions at the end of Stage 3 of $\eta^{JNU}$.
\end{claim}

\begin{proof}[Proof of Claim 5]
   Let $D'(A)=C'(A)\setminus B_0$. Since $B_0\subseteq C'(A)$, we have $C'(A)=B_0\sqcup D'(A)$. We want to show that $|D'(A)|\leq\kappa$.

By Claim 1, exactly $q^o$ applicants are assigned original open-category positions under $\eta'$.

For each reserve category
$r\in R\setminus\{OBC\}$, the number
of applicants assigned category-$r$
positions under $\eta'$ cannot exceed
either the capacity $q^r$ or the number
of category-$r$ applicants remaining
after the original open-category
assignments. Hence,
\[
\left|\left\{i\in A:
(i,r)\in\eta'(A,T)\right\}\right|
\leq\min\{q^r,|A^{1,r}|\}.
\]

Furthermore, by the combined capacity
constraint in Definition~9,
\[
\left|\left\{(i,\tau)\in\eta'(A,T):
\tau\in\{d,OBC\}\right\}\right|
\leq q^{OBC}.
\]

Since every applicant is assigned at
most one position, it follows that
\[
|C'(A)|
\leq
q^o+
\sum_{r\in R\setminus\{OBC\}}
\min\{q^r,|A^{1,r}|\}
+q^{OBC}.
\]

On the other hand, the sets constituting
$B_0$ are pairwise disjoint. Therefore,
\[
|B_0|=
q^o+
\sum_{r\in R\setminus\{OBC\}}
\min\{q^r,|A^{1,r}|\}
+m.
\]

Combining these observations yields
\[
\begin{aligned}
|D'(A)|
&=|C'(A)|-|B_0| \leq q^{OBC}-m = \kappa.
\end{aligned}
\]

Thus, any assignment rule satisfying
the four axioms can select at most
$\kappa$ applicants in addition to
the members of $B_0$. 
\end{proof}

\begin{claim}
    $\eta^{JNU}$ merit dominates $\eta^{'}$.
\end{claim}

\begin{proof}[Proof of Claim 6]
Let $R(A)=A\setminus B_0$ denote the set of applicants remaining
after the first three stages of $\eta^{JNU}$.

By construction, Stage 4 of $\eta^{JNU}$
selects the highest-scoring
$\min\{\kappa,|R(A)|\}$ applicants
from $R(A)$. Let $D^{JNU}(A)$ denote
this set. Thus,
\[
C^{JNU}(A)=B_0\sqcup D^{JNU}(A).
\]

By Claims 4 and 5, $C'(A)=B_0\sqcup D'(A)$, where $D'(A)\subseteq R(A)$ and $|D'(A)|\leq\kappa$.

If $|R(A)|\leq\kappa$, then
$\eta^{JNU}$ selects every applicant
in $R(A)$. Consequently,
\[
C^{JNU}(A)=A
\]
and hence
\[
C'(A)\subseteq C^{JNU}(A).
\]
It follows that
$C^{JNU}(A)\succeq_M C'(A)$.

Now suppose that $|R(A)|>\kappa$.
In this case, $D^{JNU}(A)$ consists
of the $\kappa$ highest-scoring
applicants in $R(A)$.

Let $s=|D'(A)|\leq\kappa$, and order
the members of $D^{JNU}(A)$ and
$D'(A)$ according to their merit
rankings:
\[
i_1\succ i_2\succ\cdots\succ i_\kappa,
\]
and
\[
j_1\succ j_2\succ\cdots\succ j_s,
\]
respectively.

Since $D^{JNU}(A)$ consists of the
$\kappa$ highest-scoring applicants
in $R(A)$ and $D'(A)$ is an
$s$-element subset of $R(A)$,
we have $\theta(i_\ell)\geq\theta(j_\ell)$ for every $\ell=1,\ldots,s$.

Define a mapping $g:C'(A)\longrightarrow C^{JNU}(A)$
by
\[
g(i)=
\begin{cases}
i, & \text{if }i\in B_0,\\
i_\ell, & \text{if }i=j_\ell
          \in D'(A),\quad \ell=1,\ldots,s.
\end{cases}
\]

Since $B_0$ and $R(A)$ are disjoint,
the mapping $g$ is well-defined and
injective. Moreover, $\theta(g(i))\geq\theta(i)$ for every $i\in C'(A)$.
\end{proof}

We next show that $C^{JNU}(A)\succeq_M C'(A)$.

First,
\[
\begin{aligned}
|C^{JNU}(A)|
&=|B_0|+\kappa \geq |B_0|+s =|C'(A)|.
\end{aligned}
\]

Second, fix any $k\leq|C'(A)|$.
Let $Y_k$ denote the set of the
$k$ highest-scoring applicants in
$C'(A)$, and let $j^k$ denote the
lowest-scoring applicant in $Y_k$.

Since $g$ is injective, $g(Y_k)$
consists of $k$ distinct applicants
in $C^{JNU}(A)$. Furthermore,
\[
\theta(g(i))\geq\theta(i)
\geq\theta(j^k)
\qquad\text{for every }i\in Y_k.
\]

Therefore, $C^{JNU}(A)$ contains
at least $k$ applicants whose merit
scores are weakly higher than the
merit score of the $k$th
highest-scoring applicant in $C'(A)$.

It follows that the merit score of
the $k$th highest-scoring applicant
in $C^{JNU}(A)$ is weakly higher
than that of the $k$th highest-scoring
applicant in $C'(A)$.

Since this holds for every
$k\leq|C'(A)|$, we conclude that
\[
C^{JNU}(A)\succeq_M C'(A).
\]

Since $A$, $T$, and $\eta'$ were
arbitrary, it follows that $\eta^{JNU}\succeq_M\eta'$ for every assignment rule $\eta'$ satisfying the OA$^{*}$ principle, WCF$^{*}$, mQFE, and OBC de-reservation$^{*}$.

We next show that every merit-optimal rule in $\mathbb{A}^*$ induces the choice rule $C^{JNU}$.

Let $\eta \in \mathbb{A}^*$ be merit-optimal in $\mathbb{A}^*$, and
fix $A \subseteq I$. Since $\eta^{JNU} \in \mathbb{A}^*$, merit-optimality of
$\eta$ gives $C^{\eta}(A) \succeq^M C^{JNU}(A)$. By the argument above,
$C^{JNU}(A) \succeq_M C^{\eta}(A)$. The two relations imply
$|C^{\eta}(A)| = |C^{JNU}(A)|$ and, for every $k \leq |C^{JNU}(A)|$, that the
$k$th highest-scoring applicant in $C^{\eta}(A)$ and the $k$th highest-scoring
applicant in $C^{JNU}(A)$ have the same merit score. Since merit scores are
distinct, these two applicants coincide for every $k$. Hence
$C^{\eta}(A) = C^{JNU}(A)$. Since $A$ was arbitrary, $C^{\eta} = C^{JNU}$.

\section*{Proof of Proposition \ref{prop:etaJNU}}

Fix a category membership profile $T\in\mathcal{T}$.
For every nonempty set of applicants $A\subseteq I$, let $C^o(A)= \left\{i\in A:\operatorname{rank}_A(i)\leq q^o\right\}$ denote the set of applicants assigned original open-category positions under $\eta^{JNU}$. Let
\[
\theta^o(A)=\min\{\theta(i):i\in C^o(A)\}
\]
denote the original open-category cutoff, and let $A^1=A\setminus C^o(A)$ be the set of applicants remaining after Stage 1.

For each $r\in R\setminus\{OBC\}$, let $A^{1,r}=\{i\in A^1:t(i)=r\}$, and let $C^r(A^1)$ denote the set of the $\min\{q^r,|A^{1,r}|\}$ highest-scoring applicants in $A^{1,r}$.

Define
\[
B_0(A)=C^o(A)\cup
\bigcup_{r\in R\setminus\{OBC\}}C^r(A^1).
\]
Thus, $B_0(A)$ is the set of applicants selected in
the first two stages of $\eta^{JNU}$. Let
\[
A^2=A\setminus B_0(A)
\]
denote the set of applicants remaining after these stages.

The set of remaining eligible OBC applicants is
\[
E^{OBC}(A)=
\left\{
i\in A^2:t(i)=OBC,\ 
\theta(i)\geq\theta^o(A)-\lambda
\right\}.
\]
Let $m(A)=\min\{q^{OBC},|E^{OBC}(A)|\}$, and let $C^{OBC}(A^2)$ denote the set of the $m(A)$ highest-scoring applicants in $E^{OBC}(A)$.

The number of vacant OBC positions at the end of
Stage 3 is
\[
\kappa(A)=q^{OBC}-m(A).
\]

Finally, let $R(A)=A^2\setminus C^{OBC}(A^2)$ denote the set of applicants remaining after Stage 3, and let $C^D(A)$ denote the set of the $\min\{\kappa(A),|R(A)|\}$ highest-scoring applicants in $R(A)$.

By construction,
\[
C^{JNU}(A)=
B_0(A)\sqcup C^{OBC}(A^2)\sqcup C^D(A).
\]
For the empty applicant set, define
$C^{JNU}(\varnothing)=B_0(\varnothing)=\varnothing$.

We first show that $\eta^{JNU}$ satisfies consistency. Fix $A\subseteq I$ and $i\in I\setminus A$. Let $B=A\cup\{i\}$ and suppose that $i\notin C^{JNU}(B)$. We want to show that $C^{JNU}(A)=C^{JNU}(B)$.

Since $i$ is not selected under $C^{JNU}(B)$,
she is not selected in any of the four stages of
$\eta^{JNU}(B,T)$.

First, since $i$ is not assigned an original
open-category position in Stage 1, removing $i$
from $B$ does not change the set of applicants
assigned original open-category positions.
Hence, $C^o(A)=C^o(B)$. In particular, the original open-category cutoff
remains unchanged. That is, $\theta^o(A)=\theta^o(B)$. Let $B^1=B\setminus C^o(B)$. Since the original open-category selections are
identical under the two applicant sets, we have
\[
A^1=B^1\setminus\{i\}.
\]

Next, consider Stage 2. For every reserve category
$r\in R\setminus\{OBC\}$, if $t(i)\neq r$, removing
$i$ does not change the set of applicants considered
for category-$r$ positions.

If $t(i)=r$, then, since $i$ is not selected in
Stage 2 under $\eta^{JNU}(B,T)$, all category-$r$
positions must be assigned to applicants with
higher priority than $i$. Removing $i$ therefore
does not affect the category-$r$ selection.

Thus,  $C^r(A^1)=C^r(B^1)$ for every $r\in R\setminus\{OBC\}$. Together with the equality of the original open-category selections, this implies
\[
B_0(A)=B_0(B).
\]

Let $B^2=B\setminus B_0(B)$. It follows that $A^2=B^2\setminus\{i\}$.

Now consider Stage 3. Since the original open-category
cutoff is unchanged, the OBC eligibility threshold
is also unchanged. Therefore,
\[
E^{OBC}(A)=E^{OBC}(B)\setminus\{i\}.
\]

If $i\notin E^{OBC}(B)$, the set of eligible OBC
applicants is identical under the two applicant sets.

If $i\in E^{OBC}(B)$, then, since $i$ is not selected
in Stage 3 under $\eta^{JNU}(B,T)$, all OBC positions
must be assigned to eligible OBC applicants with
higher priority than $i$. Removing $i$ does not
affect the set of applicants assigned OBC positions.

Thus, in either case, $C^{OBC}(A^2)=C^{OBC}(B^2)$. In particular, $m(A)=m(B)$, and hence $\kappa(A)=\kappa(B)$.

Therefore, the number of positions transferred to category $D$ is unchanged when $i$ is removed. Moreover, the sets of applicants remaining at the beginning of Stage 4 satisfy $R(A)=R(B)\setminus\{i\}$.

Since $i$ is not selected in Stage 4 under
$\eta^{JNU}(B,T)$, removing $i$ does not affect
the selection of the highest-scoring applicants
from the remaining pool, given the unchanged
capacity $\kappa(A)=\kappa(B)$.

Consequently, $C^D(A)=C^D(B)$.

Combining the equality of selections at all four
stages, we obtain
\[
\begin{aligned}
C^{JNU}(A)
&=B_0(A)\sqcup C^{OBC}(A^2)\sqcup C^D(A)\\
&=B_0(B)\sqcup C^{OBC}(B^2)\sqcup C^D(B) = C^{JNU}(B).
\end{aligned}
\]

Since $A$ and $i$ were arbitrary, $C^{JNU}$
satisfies consistency.

We next show that $\eta^{JNU}$ satisfies size monotonicity. Fix $A\subseteq I$ and $i\in I\setminus A$, and let $B=A\cup\{i\}$. We show that $|C^{JNU}(A)|\leq |C^{JNU}(B)|$.

We first establish a cardinality identity for
$C^{JNU}$.

For every nonempty applicant set $A\subseteq I$,
the number of applicants selected in Stage 3 is
$m(A)$. The number of applicants remaining at the
beginning of Stage 4 is
\[
|R(A)|=|A^2|-m(A).
\]
Since the number of de-reserved positions is
$\kappa(A)=q^{OBC}-m(A)$, the number of applicants
selected in Stage 4 is
\[
|C^D(A)|
=
\min\{q^{OBC}-m(A),|A^2|-m(A)\}.
\]

Therefore, the total number of applicants selected
in Stages 3 and 4 is
\[
\begin{aligned}
|C^{OBC}(A^2)|+|C^D(A)|
&=m(A)+
\min\{q^{OBC}-m(A),|A^2|-m(A)\}\\
&=\min\{q^{OBC},|A^2|\}.
\end{aligned}
\]

This identity establishes that the total number of
applicants selected in Stages 3 and 4 is independent
of the endogenous OBC eligibility threshold,
conditional on the set of applicants remaining
after Stage 2.

Since $A^2=A\setminus B_0(A)$, we obtain
\[
\begin{aligned}
|C^{JNU}(A)|
&=|B_0(A)|+\min\{q^{OBC},|A^2|\}\\
&=|B_0(A)|+
\min\{q^{OBC},|A|-|B_0(A)|\}\\
&=\min\{|A|,|B_0(A)|+q^{OBC}\}.
\end{aligned}
\]
The identity also holds trivially for $A=\varnothing$.
Thus, for every $A\subseteq I$,
\begin{equation}\label{eq:JNU_cardinality}
|C^{JNU}(A)|
=
\min\{|A|,|B_0(A)|+q^{OBC}\}.
\end{equation}

We next show that $|B_0(A)|\leq |B_0(B)|$. Recall that
\[
|B_0(A)|
=
\min\{|A|,q^o\}
+
\sum_{r\in R\setminus\{OBC\}}
\min\{q^r,|A^{1,r}|\}.
\]

Since $A\subseteq B$, we have $|C^o(A)|=\min\{|A|,q^o\} \leq \min\{|B|,q^o\}=|C^o(B)|$.

Moreover, we claim that $A\setminus C^o(A) \subseteq B\setminus C^o(B)$.

To see this, consider any applicant
$j\in A\setminus C^o(A)$. Since $j$ is not
among the $q^o$ highest-scoring applicants
in $A$, there are at least $q^o$ applicants
in $A$ with higher priority than $j$.
Since $A\subseteq B$, all of these applicants
also belong to $B$. Hence, there are at least
$q^o$ applicants in $B$ with higher priority
than $j$, implying that $j\notin C^o(B)$.

Thus, $A^1\subseteq B^1$, where $B^1=B\setminus C^o(B)$. 

For each reserve category $r\in R\setminus\{OBC\}$, define $B^{1,r}=\{j\in B^1:t(j)=r\}$. Since $A^1\subseteq B^1$, it follows that $A^{1,r}\subseteq B^{1,r}$ for every $r\in R\setminus\{OBC\}$.

Thus, $\min\{q^r,|A^{1,r}|\} \leq \min\{q^r,|B^{1,r}|\}$ for every $r\in R\setminus\{OBC\}$. Combining these inequalities, we obtain
\[
\begin{aligned}
|B_0(A)|
&=\min\{|A|,q^o\}
+\sum_{r\in R\setminus\{OBC\}}
\min\{q^r,|A^{1,r}|\}\\
&\leq
\min\{|B|,q^o\}
+\sum_{r\in R\setminus\{OBC\}}
\min\{q^r,|B^{1,r}|\} = |B_0(B)|.
\end{aligned}
\]

Finally, since $|A|\leq |B|$ and
$|B_0(A)|\leq |B_0(B)|$, equation
\eqref{eq:JNU_cardinality} implies
\[
\begin{aligned}
|C^{JNU}(A)|
&=\min\{|A|,|B_0(A)|+q^{OBC}\}\\
&\leq
\min\{|B|,|B_0(B)|+q^{OBC}\} = |C^{JNU}(B)|.
\end{aligned}
\]

Since $A$ and $i$ were arbitrary, $C^{JNU}$ satisfies size monotonicity. 

We have shown the violation of the substitutes condition in the main text. 

\section*{Proof of Theorem 4}

We first prove a lemma that says that, at each step, the set of applicants selected under $\eta^x$ is contained in the set selected under $\eta^y$ up to that step. In particular, for every $r\in\{o\}\cup R$, the number of type-$r$ applicants selected through Step $k$ under $\eta^y$ is weakly greater than under $\eta^x$.

\begin{lemma}\label{lem:sequential-monotonicity}
Fix $A \subseteq I$, $T \in \mathcal{T}$, and the expanded precedence order
\[
  o - T_1 - c_1 - T_2 - c_2 - T_3 - c_3 - T_4 - c_4 - T_5 .
\]
Let $x = (x_1,\ldots,x_{10}) \in \mathbb{Z}_+^{10}$ and $y = (y_1,\ldots,y_{10}) \in \mathbb{Z}_+^{10}$
be two capacity vectors, and let $K \in \{1,\ldots,10\}$. Suppose that
\begin{enumerate}
  \item[(i)] $x_1 = y_1$,
  \item[(ii)] $x_{2j+1} = y_{2j+1}$ for every $j \in \{1,\ldots,4\}$ such that $2j+1 \leq K$, and
  \item[(iii)] $x_{2j} \leq y_{2j}$ for every $j \in \{1,\ldots,5\}$ such that $2j \leq K$.
\end{enumerate}
For $z \in \{x,y\}$, let $A^z_0 = A$, and let $A^z_k$ denote the set of applicants remaining
after Step $k$ of $\eta^z(A,T)$, for $k = 1,\ldots,10$. Then, for every $k \in \{0,1,\ldots,K\}$,
\[
  A^y_k \subseteq A^x_k .
\]
\end{lemma}

\begin{proof}
We prove $A^y_k \subseteq A^x_k$ for every $k \in \{0,1,\ldots,K\}$ by induction on $k$.
The claim holds for $k = 0$ because $A^x_0 = A^y_0 = A$.

Fix $k \in \{1,\ldots,K\}$ and suppose $A^y_{k-1} \subseteq A^x_{k-1}$. For $z \in \{x,y\}$,
let $S^z_k = A^z_{k-1} \setminus A^z_k$ be the set of applicants selected at Step $k$ under
$\eta^z$. We first show that
\[
  S^x_k \cap A^y_{k-1} \subseteq S^y_k .
\]

If Step $k$ is an open-category step (the original $o$ step or one of $T_1,\ldots,T_5$),
it selects the highest-ranked remaining applicants according to $\succ$, up to its
capacity. Moreover, $x_k \leq y_k$: this follows from (i) if $k = 1$ and from (iii) if
Step $k$ is one of $T_1,\ldots,T_5$. Let $i \in S^x_k \cap A^y_{k-1}$. Then fewer than $x_k$
applicants in $A^x_{k-1}$ are ranked above $i$. Since $A^y_{k-1} \subseteq A^x_{k-1}$ and
$x_k \leq y_k$, fewer than $y_k$ applicants in $A^y_{k-1}$ are ranked above $i$. Thus
$i \in S^y_k$.

If Step $k$ is a reserved-category step $c_j \in R$, then $x_k = y_k$ by (ii). Let
$i \in S^x_k \cap A^y_{k-1}$. Then $t(i) = c_j$ (and, when $c_j = OBC$, $i$ satisfies the
fixed eligibility threshold $\theta^{OBC}$), and fewer than $x_k$ relevant applicants
for category $c_j$ in $A^x_{k-1}$ are ranked above $i$. Since
$A^y_{k-1} \subseteq A^x_{k-1}$ and $x_k = y_k$, fewer than $y_k$ relevant applicants for
category $c_j$ in $A^y_{k-1}$ are ranked above $i$. Hence $i \in S^y_k$.

Thus, in either case, $S^x_k \cap A^y_{k-1} \subseteq S^y_k$. Now let $i \in A^y_k$. Then
$i \in A^y_{k-1} \subseteq A^x_{k-1}$ and $i \notin S^y_k$. If $i \in S^x_k$, then
$i \in S^x_k \cap A^y_{k-1} \subseteq S^y_k$, a contradiction. Hence $i \notin S^x_k$, so
$i \in A^x_{k-1} \setminus S^x_k = A^x_k$. Therefore,
\[
  A^y_k \subseteq A^x_k .
\]
This completes the induction.
\end{proof}

\begin{proof}[Proof of Theorem 4]

Fix an arbitrary set of applicants $A \subseteq I$, a category profile $T \in \mathcal{T}$,
and a transfer scheme $f \in \mathcal{F}$. Let $C^{FT}$ and $C^{f}$ denote the choice rules
induced by $\eta^{FT}$ and $\eta^{f}$, respectively.

We want to show that $C^{FT}(A) \succeq^M C^{f}(A)$ or $C^{f}(A) \subseteq C^{FT}(A)$. If
$|A| \leq q^{o}$, both rules select all applicants. Hence, suppose $|A| > q^{o}$.

Let $C^{o}(A) = \{i \in A : rank_A(i) \leq q^{o}\}$ and $A^1 = A \setminus C^{o}(A)$. Since
the original open-category positions are processed first, $C^{o}(A)$ is selected under
every assignment rule in $\mathbb{A}$.

For each $r \in R \setminus \{OBC\}$, let $A^1_r = \{i \in A^1 : t(i) = r\}$ and let
$C^{r}(A^1)$ be the set of the $\min\{|A^1_r|, q^{r}\}$ highest-scoring applicants in $A^1_r$.

Define $A^1_{OBC} = \{i \in A^1 : t(i) = OBC,\ \theta(i) \geq \theta^{OBC}\}$ and let
$m^{FT} = \min\{|A^1_{OBC}|, q^{OBC}\}$ and $\kappa^{FT} = q^{OBC} - m^{FT}$. Let
$C^{OBC}(A^1)$ be the set of the $m^{FT}$ highest-scoring applicants in $A^1_{OBC}$.

Under $\eta^{FT}$, all de-reserved positions are processed at $T_5$. Define
\[
  B_0 = C^{o}(A) \cup \bigcup_{r \in R \setminus \{OBC\}} C^{r}(A^1) \cup C^{OBC}(A^1),
\]
and
\[
  R^{FT}(A) = A \setminus B_0 .
\]
Let $C^{T_5}(A)$ consist of the $\min\{\kappa^{FT}, |R^{FT}(A)|\}$ highest-scoring applicants
in $R^{FT}(A)$. Then
\[
  C^{FT}(A) = B_0 \cup C^{T_5}(A).
\]
Note that $C^{FT}(A)$ does not depend on the precedence order $\sigma$ of the reserved
categories: under $\eta^{FT}$ no de-reserved open-category position is processed before
a reserved category, and the reserved categories draw from pairwise disjoint sets of
applicants. Hence, the argument below applies regardless of the precedence order
$\sigma$ underlying $\eta^{f}$.

We will first establish that $B_0 \subseteq C^{f}(A)$.

The iterative procedure defining $\eta^{f}(A,T)$ terminates. Indeed, $0 \leq \kappa_n \leq
x^{n}_{r^*}$ in every iteration $n$, and whenever $\kappa_n > 0$ we have
$x^{n+1}_{r^*} = x^{n}_{r^*} - \kappa_n < x^{n}_{r^*}$. Since $x^{1}_{r^*} = q^{OBC}$ and the
OBC capacity is a nonnegative integer, $\kappa_n > 0$ can hold in at most $q^{OBC}$
iterations. Let $N$ denote the terminal iteration of $\eta^{f}(A,T)$, with terminal
capacity vector $x^{N} = (x^{N}_1,\ldots,x^{N}_{10})$. Then
$\eta^{f}(A,T) = \eta^{x^{N}}(A,T)$ and $\kappa_N = 0$.

Consider any $r \in R \setminus \{OBC\}$. Every applicant $i \in C^{r}(A^1)$ is either
selected at a de-reserved open-category stage preceding category $r$, or remains
available when category $r$ is processed. In the latter case, fewer than $q^{r}$
remaining applicants of type $r$ have higher merit than $i$. Hence, the category-$r$
stage selects $i$. Therefore,
\[
  C^{r}(A^1) \subseteq C^{f}(A) \qquad \text{for every } r \in R \setminus \{OBC\}.
\]

Next, we consider the OBC category. If $|A^1_{OBC}| \geq q^{OBC}$, all OBC positions are
filled in Iteration 1. No de-reservation occurs, so $\eta^{f}(A,T) = \eta^{FT}(A,T)$.

Suppose instead that $e \equiv |A^1_{OBC}| < q^{OBC}$. For each iteration $n$, let $b_n$ be
the number of applicants in $A^1_{OBC}$ selected before the OBC stage, excluding those
selected at the original open-category stage.

Since $A^1_{OBC} \subseteq A^1$, these are precisely the eligible OBC applicants selected
at the de-reserved open-category stages preceding OBC.

Let $r^*$ denote the step of the OBC category in the expanded precedence order.
By Definition~\ref{def:transferscheme}, $x^{n+1}_{2j} =f_{j}(v_{n+1})  \ge f_{j}(v_n)=x^{n}_{2j}$ for
$j = 1,\ldots,5$, while the capacities of the original open-category step and of all
non-OBC reserved-category steps are the same in $x^{n}$ and $x^{n+1}$. Hence $x = x^{n}$
and $y = x^{n+1}$ satisfy conditions (i)--(iii) of Lemma~\ref{lem:sequential-monotonicity} with $K = r^* - 1$, and
therefore
\[
  A^{x^{n+1}}_k \subseteq A^{x^{n}}_k \qquad \text{for every } k \leq r^* - 1 .
\]
In particular, $b_{n+1} \geq b_n$.

Exactly $e - b_n$ eligible OBC applicants remain available at the OBC stage in
iteration $n$. Thus, we have
\[
  \kappa_n = \max\{0,\ x^{n}_{r^*} - (e - b_n)\},
\]
and the capacity update gives
\[
  x^{n+1}_{r^*} = x^{n}_{r^*} - \kappa_n = \min\{x^{n}_{r^*},\ e - b_n\}.
\]

In Iteration 1, all de-reserved open-category capacities are zero. Thus, $b_1 = 0$ and
$x^{1}_{r^*} = q^{OBC} > e$. So, we have
\[
  \kappa_1 = q^{OBC} - e > 0, \qquad x^{2}_{r^*} = e .
\]

If $N = 2$, termination implies $0 = \kappa_2 = \max\{0,\ e - (e - b_2)\} = b_2$. Hence, all
$e$ eligible OBC applicants are selected at the OBC stage.

If $N > 2$, then $\kappa_{N-1} > 0$. So, $x^{N}_{r^*} = e - b_{N-1}$. Since
$b_N \geq b_{N-1}$, $e - b_N \leq x^{N}_{r^*}$. On the other hand, $\kappa_N = 0$ implies
$x^{N}_{r^*} \leq e - b_N$. Therefore, $x^{N}_{r^*} = e - b_N$.

Thus, in the final iteration, the $b_N$ eligible OBC applicants are selected before the
OBC stage, and the remaining $e - b_N$ are selected at the OBC stage.

It follows that every applicant in $A^1_{OBC}$ is selected in $\eta^{f}$. In particular,
\[
  C^{OBC}(A^1) \subseteq C^{f}(A).
\]

Combining these inclusions with the common original open-category selection
establishes $B_0 \subseteq C^{f}(A)$.

We next establish merit domination.

Define $D^{f}(A) = C^{f}(A) \setminus B_0$. Since $B_0 \subseteq C^{f}(A)$,
$C^{f}(A) = B_0 \cup D^{f}(A)$.

Under $\eta^{f}$, the original open-category stage selects exactly $q^{o}$ applicants.
Each reserved category $r \in R \setminus \{OBC\}$ selects at most
$\min\{|A^1_r|, q^{r}\}$ applicants.

Moreover, the combined capacity of the filled OBC positions and the de-reserved
open-category positions is $q^{OBC}$. Therefore,
\[
  |C^{f}(A)| \leq q^{o} + \sum_{r \in R \setminus \{OBC\}} \min\{|A^1_r|, q^{r}\} + q^{OBC}.
\]

Since $|B_0| = q^{o} + \sum_{r \in R \setminus \{OBC\}} \min\{|A^1_r|, q^{r}\} + m^{FT}$, we
obtain
\[
  |D^{f}(A)| \leq q^{OBC} - m^{FT} = \kappa^{FT}.
\]

Furthermore, $D^{f}(A) \subseteq R^{FT}(A)$.

If $|R^{FT}(A)| < \kappa^{FT}$, the forward-transfer rule selects every applicant in
$R^{FT}(A)$. Hence, $C^{FT}(A) = A$, and therefore $C^{f}(A) \subseteq C^{FT}(A)$.

Now suppose that $|R^{FT}(A)| \geq \kappa^{FT}$. Then $C^{T_5}(A)$ consists of the
$\kappa^{FT}$ highest-scoring applicants in $R^{FT}(A)$.

Let $s = |D^{f}(A)| \leq \kappa^{FT}$. Order the applicants in $D^{f}(A)$ and $C^{T_5}(A)$
according to merit: $j_1 \succ \cdots \succ j_s$ and
$i_1 \succ \cdots \succ i_{\kappa^{FT}}$.

Since $D^{f}(A)$ is an $s$-element subset of $R^{FT}(A)$, while $C^{T_5}(A)$ contains its
$\kappa^{FT}$ highest-scoring applicants, we have
\[
  \theta(i_\ell) \geq \theta(j_\ell), \qquad \ell = 1,\ldots,s .
\]

Define an injection $g : C^{f}(A) \longrightarrow C^{FT}(A)$ by $g(i) = i$ for every
$i \in B_0$ and $g(j_\ell) = i_\ell$ for $\ell = 1,\ldots,s$.

Because $B_0$ and $R^{FT}(A)$ are disjoint, $g$ is well-defined and injective. Moreover,
\[
  \theta(g(i)) \geq \theta(i) \qquad \text{for every } i \in C^{f}(A).
\]

We now verify that $C^{FT}(A) \succeq^M C^{f}(A)$. First, since $B_0$ and $R^{FT}(A)$ are
disjoint,
\[
  |C^{FT}(A)| = |B_0| + \kappa^{FT} \geq |B_0| + s = |C^{f}(A)| .
\]
Second, fix $k \leq |C^{f}(A)|$, let $Y_k$ be the set of the $k$ highest-scoring applicants
in $C^{f}(A)$, and let $i^{k}$ be the lowest-scoring applicant in $Y_k$, that is, the $k$th
highest-scoring applicant in $C^{f}(A)$. Since $g$ is injective, $g(Y_k)$ consists of $k$
distinct applicants in $C^{FT}(A)$, and for every $i \in Y_k$ we have
$\theta(g(i)) \geq \theta(i) \geq \theta(i^{k})$. Hence, $C^{FT}(A)$ contains at least $k$
applicants whose scores are weakly higher than $\theta(i^{k})$, so the $k$th
highest-scoring applicant in $C^{FT}(A)$ has a weakly higher score than the $k$th
highest-scoring applicant in $C^{f}(A)$.

Thus, $C^{FT}(A) \succeq^M C^{f}(A)$. We have shown that, for every applicant set $A$,
$C^{FT}(A) \succeq^M C^{f}(A)$ or $C^{f}(A) \subseteq C^{FT}(A)$.

Since $A$, $T$, and $f \in \mathcal{F}$ were arbitrary, $\eta^{FT} \succeq^M \eta^{f}$ for
every $f \in \mathcal{F}$.

\end{proof}

\section*{Proof of Theorem 5}

Throughout, fix $T \in \mathcal{T}$, the expanded precedence order, and a transfer scheme
$f \in \mathcal{F}$. Let $r^*$ denote the step of the OBC category. For
$v \in \{0,\ldots,q^{OBC}\}$, let $x(v) \in \mathbb{Z}_+^{10}$ be the capacity vector with
\[
  x_1(v) = q^{o}, \qquad x_{2j}(v) = f_j(v) \ \ (j = 1,\ldots,5), \qquad
  x_{r^*}(v) = q^{OBC} - v,
\]
and $x_{2j+1}(v) = q^{c_j}$ for every $j$ with $c_j \neq OBC$. In iteration $n$ of the
procedure defining $\eta^{f}$, the capacity vector is $x^n = x(v_n)$, where $v_1 = 0$ and
$v_{n+1} = v_n + \kappa_n$.

For each step $k$, let $E_k \subseteq I$ be the set of applicants who are relevant at
Step $k$: $E_k = I$ if Step $k$ is an open-category step, $E_k = \{i \in I : t(i) = c_j\}$
if Step $k$ is the step of $c_j \neq OBC$, and
$E_{r^*} = \{i \in I : t(i) = OBC,\ \theta(i) \geq \theta^{OBC}\}$. For $B \subseteq I$ and
$z \in \mathbb{Z}_+^{10}$, let $B^z_0 = B$, let $B^z_k$ be the set of applicants remaining
after Step $k$ of $\eta^{z}(B,T)$, and let $S^z_k(B) = B^z_{k-1} \setminus B^z_k$ be the set
selected at Step $k$. By the definition of $\eta^{z}$, for every $k$,
\begin{equation}\label{eq:sel}
  i \in S^z_k(B) \iff i \in B^z_{k-1} \cap E_k \ \text{and}\
  \bigl|\{ j \in B^z_{k-1} \cap E_k : j \succ i \}\bigr| < z_k ,
\end{equation}
and therefore
\begin{equation}\label{eq:size}
  |S^z_k(B)| = \min\bigl\{ z_k ,\ |B^z_{k-1} \cap E_k| \bigr\}.
\end{equation}

\begin{lemma}\label{lem:B}
Let $A \subseteq B \subseteq I$, let $x, y \in \mathbb{Z}_+^{10}$, and let
$K \in \{1,\ldots,10\}$. Suppose that, for every $k \leq K$,
\begin{enumerate}
  \item[(a)] if Step $k$ is an open-category step, then $x_k \leq y_k$, and
  \item[(b)] if Step $k$ is a reserved-category step, then $x_k = y_k$ or
             $S^y_k(A) = A^y_{k-1} \cap E_k$.
\end{enumerate}
Then $A^y_k \subseteq B^x_k$ for every $k \in \{0,1,\ldots,K\}$.
\end{lemma}

\begin{proof}
By induction on $k$. For $k = 0$, $A^y_0 = A \subseteq B = B^x_0$. Fix
$k \in \{1,\ldots,K\}$ and suppose $A^y_{k-1} \subseteq B^x_{k-1}$. We first show
\begin{equation}\label{eq:key}
  S^x_k(B) \cap A^y_{k-1} \subseteq S^y_k(A).
\end{equation}
Let $i \in S^x_k(B) \cap A^y_{k-1}$. By \eqref{eq:sel}, $i \in E_k$ and fewer than $x_k$
members of $B^x_{k-1} \cap E_k$ are ranked above $i$. In particular,
$i \in A^y_{k-1} \cap E_k$. If $x_k \leq y_k$, then, since
$A^y_{k-1} \cap E_k \subseteq B^x_{k-1} \cap E_k$, fewer than $y_k$ members of
$A^y_{k-1} \cap E_k$ are ranked above $i$, so $i \in S^y_k(A)$ by \eqref{eq:sel}. By (a)
and (b), the only remaining case is that Step $k$ is a reserved-category step with
$S^y_k(A) = A^y_{k-1} \cap E_k$, and then $i \in S^y_k(A)$ as well. This proves
\eqref{eq:key}.

Now let $i \in A^y_k$. Then $i \in A^y_{k-1} \subseteq B^x_{k-1}$ and $i \notin S^y_k(A)$.
If $i \in S^x_k(B)$, then $i \in S^y_k(A)$ by \eqref{eq:key}, a contradiction. Hence
$i \in B^x_{k-1} \setminus S^x_k(B) = B^x_k$.
\end{proof}

\begin{lemma}\label{lem:D}
Let $A \subseteq I$, $i \in I \setminus A$, $B = A \cup \{i\}$, and
$z \in \mathbb{Z}_+^{10}$. Then, for every $k \in \{0,1,\ldots,10\}$,
$A^z_k \subseteq B^z_k$ and $|B^z_k \setminus A^z_k| \leq 1$.
\end{lemma}

\begin{proof}

The inclusion $A_k^z\subseteq B_k^z$ follows from Lemma~\ref{lem:B} with $x=y=z$ and $K=10$. For the cardinality bound, note that $|B_0^z\setminus A_0^z|=1$.

For every $k\geq 1$, the inclusions
$A_{k-1}^z\subseteq B_{k-1}^z$ and
$A_k^z\subseteq B_k^z$ imply

\[
\begin{aligned}
|B_k^z\setminus A_k^z|
&=|B_k^z|-|A_k^z|\\
&=\bigl(|B_{k-1}^z|-|S_k^z(B)|\bigr)
-\bigl(|A_{k-1}^z|-|S_k^z(A)|\bigr)\\
&=|B_{k-1}^z\setminus A_{k-1}^z|
-\bigl(|S_k^z(B)|-|S_k^z(A)|\bigr).
\end{aligned}
\]

Since
\[
A_{k-1}^z\cap E_k\subseteq B_{k-1}^z\cap E_k,
\]
the definition of sequential selection implies
\[
\begin{aligned}
|S_k^z(A)|
&=\min\{z_k,|A_{k-1}^z\cap E_k|\}\\
&\leq\min\{z_k,|B_{k-1}^z\cap E_k|\}=|S_k^z(B)|.
\end{aligned}
\]

Thus,
\[
|B_k^z\setminus A_k^z|
\leq
|B_{k-1}^z\setminus A_{k-1}^z|.
\]

Thus, $|B_k^z\setminus A_k^z|$ is nonincreasing
in $k$. Since $|B_0^z\setminus A_0^z|=1$, we obtain
\[
|B_k^z\setminus A_k^z|\leq 1
\]
for every $k\in\{0,1,\ldots,10\}$.
\end{proof}

For $B \subseteq I$ and $v \in \{0,\ldots,q^{OBC}\}$, let $n_B(v) = \bigl| B^{x(v)}_{r^*-1} \cap E_{r^*} \bigr|$ be the number of eligible OBC applicants available at the OBC step of
$\eta^{x(v)}(B,T)$. By \eqref{eq:size}, the OBC step selects
$\min\{q^{OBC} - v,\ n_B(v)\}$ applicants, so the number of vacant OBC positions in
$\eta^{x(v)}(B,T)$ is
\begin{equation}\label{eq:vac}
  \kappa_B(v) = \max\bigl\{ 0,\ q^{OBC} - v - n_B(v) \bigr\}.
\end{equation}

\begin{lemma}\label{lem:C}
Let $B \subseteq I$.
\begin{enumerate}
  \item[(a)] $n_B(v+1) \leq n_B(v)$ for every $v \in \{0,\ldots,q^{OBC}-1\}$.
  \item[(b)] The procedure defining $\eta^{f}(B,T)$ terminates, and its terminal cumulative number of vacancies, denoted $v^*(B)$, is the smallest $v \in \{0,\ldots,q^{OBC}\}$ such that $\kappa_B(v) = 0$. Moreover, $\eta^{f}(B,T) = \eta^{x(v^*(B))}(B,T)$.
  \item[(c)] The OBC step of $\eta^{x(v^*(B))}(B,T)$ selects exactly
             $q^{OBC} - v^*(B)$ applicants. If $v^*(B) > 0$, then
             $n_B(v^*(B)) = q^{OBC} - v^*(B)$. This step selects all applicants in $B^{x(v^*(B))}_{r^*-1} \cap E_{r^*}$.
\end{enumerate}
\end{lemma}

\begin{proof}
(a) By Definition~13(2), $f_j(v) \leq f_j(v+1)$ for every $j$. The vectors $x(v)$ and
$x(v+1)$ coincide at Step 1 and at every non-OBC reserved-category step. Hence
Lemma~\ref{lem:B}, applied with $A = B$, $x = x(v)$, $y = x(v+1)$ and $K = r^* - 1$,
gives $B^{x(v+1)}_{r^*-1} \subseteq B^{x(v)}_{r^*-1}$, and so $n_B(v+1) \leq n_B(v)$.

(b) By \eqref{eq:vac}, $0 \leq \kappa_n = \kappa_B(v_n) \leq q^{OBC} - v_n$, so
$v_n \leq v_{n+1} \leq q^{OBC}$ for every $n$, with $v_{n+1} > v_n$ whenever
$\kappa_n > 0$. Since $v_n$ are integers in $\{0,\ldots,q^{OBC}\}$, there is a first
iteration $N$ with $\kappa_N = 0$. The procedure ends there, $v^*(B) = v_N$ satisfies
$\kappa_B(v_N) = 0$, and $\eta^{f}(B,T) = \eta^{x^N}(B,T) = \eta^{x(v_N)}(B,T)$.

For minimality, let $w \in \{0,\ldots,q^{OBC}\}$ satisfy $\kappa_B(w) = 0$, that is,
$n_B(w) \geq q^{OBC} - w$. We show $v_n \leq w$ for all $n \leq N$ by induction.
Clearly $v_1 = 0 \leq w$. Suppose $v_n \leq w$ and $n < N$. Then $\kappa_n > 0$, so
by \eqref{eq:vac}, $v_{n+1} = v_n + \kappa_n = q^{OBC} - n_B(v_n)$. By (a),
$n_B(v_n) \geq n_B(w)$, and hence
$v_{n+1} \leq q^{OBC} - n_B(w) \leq w$.

(c) Write $v^* = v^*(B)$. Since $\kappa_B(v^*) = 0$, \eqref{eq:vac} gives
$n_B(v^*) \geq q^{OBC} - v^*$, so the OBC step selects
$\min\{q^{OBC} - v^*, n_B(v^*)\} = q^{OBC} - v^*$ applicants. Suppose $v^* > 0$. By
(b), $\kappa_B(v^* - 1) > 0$, that is, $n_B(v^*-1) < q^{OBC} - v^* + 1$, so
$n_B(v^*-1) \leq q^{OBC} - v^*$. By (a), $n_B(v^*) \leq n_B(v^*-1) \leq q^{OBC} - v^*$.
Thus $n_B(v^*) = q^{OBC} - v^*$, and by \eqref{eq:size} the OBC step selects all
$n_B(v^*)$ applicants in $B^{x(v^*)}_{r^*-1} \cap E_{r^*}$.
\end{proof}

\begin{proof}[Proof of Theorem 5]
Fix $A \subseteq I$ and $i \in I \setminus A$, and let $B = A \cup \{i\}$. Let
$v = v^*(A)$ and $v' = v^*(B)$, and write $y = x(v)$ and $x = x(v')$. By
Lemma~\ref{lem:C}(b),
\[
  C^{f}(A) = A \setminus A^{y}_{10}, \qquad C^{f}(B) = B \setminus B^{x}_{10}.
\]
We want to show that
\begin{enumerate}
  \item[(S)] $C^{f}(B) \cap A \subseteq C^{f}(A)$,
  \item[(M)] $|C^{f}(A)| \leq |C^{f}(B)|$, and
  \item[(C)] $i \notin C^{f}(B)$ implies $C^{f}(B) = C^{f}(A)$.
\end{enumerate}
Since $A$ and $i$ are arbitrary, (M) is size monotonicity and (C) is consistency.
Property (S) is the substitutes property: given $i, j \in I$ and
$A_0 \subseteq I \setminus \{i,j\}$ with $j \notin C^{f}(A_0 \cup \{j\})$, applying (S)
to $A = A_0 \cup \{j\}$ yields $j \notin C^{f}(A_0 \cup \{i,j\})$.

\medskip
\noindent\emph{Step 1: $v' \leq v$.} By Lemma~\ref{lem:B} with both capacity vectors equal
to $x(v)$ and $K = r^*-1$, $A^{x(v)}_{r^*-1} \subseteq B^{x(v)}_{r^*-1}$, so
$n_B(v) \geq n_A(v)$. Since $\kappa_A(v) = 0$, \eqref{eq:vac} gives
$n_A(v) \geq q^{OBC} - v$. Hence $n_B(v) \geq q^{OBC} - v$, that is, $\kappa_B(v) = 0$,
and $v' \leq v$ by Lemma~\ref{lem:C}(b).

\medskip
\noindent\emph{Step 2: if $v' < v$, then $n_A(v) = q^{OBC} - v$ and
$n_B(v') = q^{OBC} - v'$.} Suppose $v' < v$. Then $v > 0$, and the first equality is
Lemma~\ref{lem:C}(c). For the second, $\kappa_B(v') = 0$ gives
$n_B(v') \geq q^{OBC} - v'$. Conversely, since $v' < v = v^*(A)$,
Lemma~\ref{lem:C}(b) gives $\kappa_A(v') > 0$, that is,
$n_A(v') \leq q^{OBC} - v' - 1$. By Lemma~\ref{lem:D} with $z = x(v')$,
$A^{x(v')}_{r^*-1} \subseteq B^{x(v')}_{r^*-1}$ and the two sets differ by at most one
applicant, so $n_B(v') \leq n_A(v') + 1 \leq q^{OBC} - v'$.

\medskip
\noindent\emph{Step 3: $A^{y}_k \subseteq B^{x}_k$ for every $k \in \{0,1,\ldots,10\}$.}
We verify the hypotheses of Lemma~\ref{lem:B} with $K = 10$. At Step 1,
$x_1 = y_1 = q^{o}$. At the step of $T_j$, $x_{2j} = f_j(v') \leq f_j(v) = y_{2j}$,
because $v' \leq v$ and each $f_j$ is nondecreasing by Definition~13(2). At every
non-OBC reserved-category step the two capacities coincide. At the OBC step, if
$v' = v$ the two capacities coincide. If $v' < v$, then by Step 2,
$n_A(v) = q^{OBC} - v = y_{r^*}$, so by \eqref{eq:size} the OBC step of $\eta^{y}(A,T)$
selects all of $A^{y}_{r^*-1} \cap E_{r^*}$, that is,
$S^{y}_{r^*}(A) = A^{y}_{r^*-1} \cap E_{r^*}$. Thus condition (b) of Lemma~\ref{lem:B}
holds in either case.

\medskip
\noindent\emph{Proof of (S).} By Step 3, $A^{y}_{10} \subseteq B^{x}_{10}$. Hence
\[
  C^{f}(B) \cap A = A \setminus B^{x}_{10} \subseteq A \setminus A^{y}_{10} = C^{f}(A).
\]

\medskip
\noindent\emph{Proof of (M).} Since the steps select disjoint sets,
\[
  |C^{f}(A)| - |C^{f}(B)| = \sum_{k=1}^{10} \bigl( |S^{y}_k(A)| - |S^{x}_k(B)| \bigr).
\]
We bound each term, using $A^{y}_{k-1} \subseteq B^{x}_{k-1}$ from Step 3.

If Step $k$ is an open-category step, let $p = |A^{y}_{k-1}|$ and $p' = |B^{x}_{k-1}|$,
so $p \leq p'$. By \eqref{eq:size}, the term equals $\min\{y_k,p\} - \min\{x_k,p'\}$.
If $x_k \leq p'$, this is at most $y_k - x_k$. If $x_k > p'$, it equals
$\min\{y_k,p\} - p' \leq p - p' \leq 0 \leq y_k - x_k$. Hence the term is at most
$y_k - x_k$, which equals $0$ for $k = 1$ and $f_j(v) - f_j(v')$ for the step of $T_j$.

If Step $k$ is a non-OBC reserved-category step, then $x_k = y_k$ and
$A^{y}_{k-1} \cap E_k \subseteq B^{x}_{k-1} \cap E_k$, so the term is at most $0$ by
\eqref{eq:size}.

If Step $k$ is the OBC step, then by Lemma~\ref{lem:C}(c),
$|S^{y}_{r^*}(A)| = q^{OBC} - v$ and $|S^{x}_{r^*}(B)| = q^{OBC} - v'$, so the term
equals $-(v - v')$.

Summing, and using $\sum_{j=1}^{5} f_j(v) = v$ and $\sum_{j=1}^{5} f_j(v') = v'$ from
Definition~13(1),
\[
  |C^{f}(A)| - |C^{f}(B)| \leq \sum_{j=1}^{5} \bigl( f_j(v) - f_j(v') \bigr) - (v - v') = 0 .
\]

\medskip
\noindent\emph{Proof of (C).} Suppose $i \notin C^{f}(B)$, so that $i \in B^{x}_k$ for
every $k$. We show by induction that, for every $k \in \{0,1,\ldots,10\}$,
\begin{equation}\label{eq:irc}
  A^{x}_k = B^{x}_k \setminus \{i\}, \qquad \text{and} \qquad
  S^{x}_k(A) = S^{x}_k(B) \ \text{ if } k \geq 1 .
\end{equation}
For $k = 0$, $A^{x}_0 = A = B \setminus \{i\}$. Fix $k \geq 1$ and suppose
$A^{x}_{k-1} = B^{x}_{k-1} \setminus \{i\}$. If $i \notin E_k$, then
$A^{x}_{k-1} \cap E_k = B^{x}_{k-1} \cap E_k$, so the same applicants are selected. If
$i \in E_k$, then $i \in B^{x}_{k-1} \cap E_k$ and $i \notin S^{x}_k(B)$, so by
\eqref{eq:sel} at least $x_k$ members of $B^{x}_{k-1} \cap E_k$ are ranked above $i$.
Hence $S^{x}_k(B)$ consists of the $x_k$ highest-ranked members of
$B^{x}_{k-1} \cap E_k$, all of whom are ranked above $i$ and therefore belong to
$A^{x}_{k-1} \cap E_k = (B^{x}_{k-1} \cap E_k) \setminus \{i\}$, where they are also the
$x_k$ highest-ranked members. Thus $S^{x}_k(A) = S^{x}_k(B)$. In both cases,
$A^{x}_k = A^{x}_{k-1} \setminus S^{x}_k(A) = (B^{x}_{k-1} \setminus \{i\}) \setminus S^{x}_k(B)
= B^{x}_k \setminus \{i\}$, which proves \eqref{eq:irc}.

By \eqref{eq:irc} at $k = r^*$, the OBC step of $\eta^{x}(A,T) = \eta^{x(v')}(A,T)$
selects the same number of applicants as the OBC step of $\eta^{x(v')}(B,T)$, which
has no vacancy. Hence $\kappa_A(v') = 0$, and Lemma~\ref{lem:C}(b) gives $v \leq v'$.
Together with Step 1, $v = v'$, so $y = x$. Therefore, using \eqref{eq:irc} at
$k = 10$ and $i \in B^{x}_{10}$,
\[
  C^{f}(A) = A \setminus A^{x}_{10}
           = (B \setminus \{i\}) \setminus (B^{x}_{10} \setminus \{i\})
           = B \setminus B^{x}_{10} = C^{f}(B).
\]
This completes the proof.
\end{proof}

\section*{Proof of Proposition \ref{prop:JNU_FT_cardinality}}
The assignment rules $\eta^{JNU}$ and $\eta^{FT}$ select the same applicants for the original open-category positions and the positions reserved for categories in $R\setminus\{OBC\}$.
Let
\[
B_0(A)=C^o(A)\cup
\bigcup_{r\in R\setminus\{OBC\}}C^r(A^1)
\]
denote this common set of selected applicants.

Under both assignment rules, any OBC position that remains vacant after the allocation of positions to eligible OBC applicants is transferred forward and
made available to all remaining applicants. Thus, irrespective of the OBC eligibility threshold, the combined number of applicants assigned
genuine OBC positions and de-reserved positions is $\min\{q^{OBC},|A\setminus B_0(A)|\}$.

It follows that $|C^{JNU}(A)| =|B_0(A)|+ \min\{q^{OBC},|A\setminus B_0(A)|\} =|C^{FT}(A)|$. Since $A$ and $T$ were arbitrary, the result follows.

\section*{Proof of Theorem 6}

Fix an applicant set $A \subseteq I$ and category profile $T = (t(i))_{i \in A}$, and suppose $\tau^{JNU}(A) \ge \tau^{PVI}$, where $\tau^{JNU}(A) = \theta^{o}(A) - \lambda$ with $\theta^{o}(A) = \min\{\theta(i) : i \in C^{o}(A)\}$ and $\tau^{PVI} = \underline{\theta}^{OBC}$. If $|A| \le q^{o}$, both rules select all of $A$ in Stage 1 / Step 1 and the cohorts coincide. Assume $|A| > q^{o}$.

In Stage 1 of $\eta^{JNU}$ and Step 1 of $\eta^{FT}$, both rules select the top $q^{o}$ applicants in $A$, forming the common set $C^{o}(A) = \{i \in A : \text{rank}_{A}(i) \le q^{o}\}$, and leave the common remaining pool $A^{1} = A \setminus C^{o}(A)$.

For each $r \in R \setminus \{OBC\}$, Stage 2 of $\eta^{JNU}$ selects the highest-scoring $\min\{|A^{1}_{r}|, q^{r}\}$ applicants from $A^{1}_{r} = \{i \in A^{1} : t(i) = r\}$, denoted $C^{r}(A^{1})$. Under $\eta^{FT}$, because $f^{FT}(\kappa) = (0, 0, 0, 0, \kappa)$ sets zero capacity for $T_{1}, \dots, T_{4}$, no open-category positions precede any reserve category $r \in R \setminus \{OBC\}$ other than $\text{Open}$ (Step 1), and the sets $A^{1}_{r}$ are disjoint across $r \in R$. Thus $\eta^{FT}$ selects the exact same set $C^{r}(A^{1})$ for every $r \in R \setminus \{OBC\}$.

Let $B_{0} = C^{o}(A) \cup \bigcup_{r \in R \setminus \{OBC\}} C^{r}(A^{1})$ and $A^{2} = A \setminus B_{0}$, with $A^{2}_{OBC} = \{i \in A^{2} : t(i) = OBC\}$.

Now let us define $E^{JNU}(A) = \{i \in A^{2}_{OBC} : \theta(i) \ge \tau^{JNU}(A)\}$ and $E^{PVI}(A) = \{i \in A^{2}_{OBC} : \theta(i) \ge \tau^{PVI}\}$. Since $\tau^{JNU}(A) \ge \tau^{PVI}$, we have $E^{JNU}(A) \subseteq E^{PVI}(A)$.

Let $m^{JNU} = \min\{q^{OBC}, |E^{JNU}(A)|\}$ and $m^{FT} = \min\{q^{OBC}, |E^{PVI}(A)|\}$ be the numbers of $OBC$ seats filled under $\eta^{JNU}$ (Stage 3) and $\eta^{FT}$ (step $r^{*}$), leaving $\kappa^{JNU} = q^{OBC} - m^{JNU}$ and $\kappa^{FT} = q^{OBC} - m^{FT}$ vacant $OBC$ seats, respectively.

Because $E^{JNU}(A) \subseteq E^{PVI}(A)$ and every applicant in $E^{JNU}(A)$ has score greater than or equal to $\tau^{JNU}(A)$ whereas every applicant in $E^{PVI}(A) \setminus E^{JNU}(A)$ has score strictly less than $\tau^{JNU}(A)$, the set $E^{JNU}(A)$ is an upper contour set of $E^{PVI}(A)$ under $\succ$. Hence, $m^{JNU} \le m^{FT}$ and $\kappa^{JNU} \ge \kappa^{FT}$. Moreover, letting $C^{OBC}_{JNU}(A)$ and $C^{OBC}_{FT}(A)$ denote the $OBC$ selections under $\eta^{JNU}$ and $\eta^{FT}$, we have $C^{OBC}_{JNU}(A) \subseteq C^{OBC}_{FT}(A)$, and $\Delta_{OBC} \equiv C^{OBC}_{FT}(A) \setminus C^{OBC}_{JNU}(A)$ has cardinality $|\Delta_{OBC}| = m^{FT} - m^{JNU} = \kappa^{JNU} - \kappa^{FT}$.

After the $OBC$ category is filled based on merit scores, the remaining pools are $R^{JNU}(A) = A^{2} \setminus C^{OBC}_{JNU}(A)$ and $R^{FT}(A) = A^{2} \setminus C^{OBC}_{FT}(A)$, satisfying $R^{JNU}(A) = R^{FT}(A) \sqcup \Delta_{OBC}$.

If $|R^{JNU}(A)| \le \kappa^{JNU}$, then $|R^{FT}(A)| = |R^{JNU}(A)| - |\Delta_{OBC}| \le \kappa^{FT}$, so both $\eta^{JNU}$ (Stage 4) and $\eta^{FT}$ (Step 10, $T_{5}$) select all remaining applicants and produce the identical cohort $C^{JNU}(A) = C^{FT}(A) = A$.

 If $|R^{JNU}(A)| > \kappa^{JNU}$, then $|R^{FT}(A)| = |R^{JNU}(A)| - (\kappa^{JNU} - \kappa^{FT}) > \kappa^{FT}$. Let $C^{D}_{JNU}(A)$ be the top $\kappa^{JNU}$ applicants of $R^{JNU}(A)$ selected in Stage 4 of $\eta^{JNU}$, and let $C^{T_{5}}_{FT}(A)$ be the top $\kappa^{FT}$ applicants of $R^{FT}(A)$ selected in Step 10 of $\eta^{FT}$. Notice that $\Delta_{OBC} \sqcup C^{T_{5}}_{FT}(A)$ is a subset of $R^{JNU}(A)$ of size $|\Delta_{OBC}| + \kappa^{FT} = \kappa^{JNU}$.

In the latter case, both $C^{JNU}(A) = B_{0} \sqcup C^{OBC}_{JNU}(A) \sqcup C^{D}_{JNU}(A)$ and $C^{FT}(A) = B_{0} \sqcup C^{OBC}_{JNU}(A) \sqcup \Delta_{OBC} \sqcup C^{T_{5}}_{FT}(A)$ have size $|B_{0}| + m^{JNU} + \kappa^{JNU} = |B_{0}| + q^{OBC}$. Ordering $C^{D}_{JNU}(A)$ as $u_{1} > \dots > u_{\kappa^{JNU}}$ and $\Delta_{OBC} \sqcup C^{T_{5}}_{FT}(A)$ as $v_{1} > \dots > v_{\kappa^{JNU}}$, since $u_{1}, \dots, u_{l}$ are the top $l$ elements of $R^{JNU}(A)$ and $v_{1}, \dots, v_{l}$ are $l$ distinct elements of $R^{JNU}(A)$, we have $\theta(u_{l}) \ge \theta(v_{l})$ for every $l = 1, \dots, \kappa^{JNU}$. Together with the common elements in $B_{0} \sqcup C^{OBC}_{JNU}(A)$, the $k$-th highest-scoring applicant in $C^{JNU}(A)$ has a weakly higher score than the $k$-th highest-scoring applicant in $C^{FT}(A)$ for all $k \le |C^{FT}(A)|$, so $C^{JNU}(A) \ge^{M} C^{FT}(A)$.

Finally, if $C^{D}_{JNU}(A) = \Delta_{OBC} \sqcup C^{T_{5}}_{FT}(A)$, then $C^{JNU}(A) = C^{FT}(A)$. If $\tau^{JNU}(A) > \tau^{PVI}$ and $C^{JNU}(A) \neq C^{FT}(A)$, then $C^{D}_{JNU}(A) \neq \Delta_{OBC} \sqcup C^{T_{5}}_{FT}(A)$. Because all merit scores $\theta$ are distinct, two unequal $\kappa^{JNU}$-element subsets of $R^{JNU}(A)$ cannot share the same ordered score vector, so $\theta(u_{l^{*}}) > \theta(v_{l^{*}})$ strictly for at least one $l^{*} \in \{1, \dots, \kappa^{JNU}\}$. Thus, $C^{JNU}(A)$ strictly merit-dominates $C^{FT}(A)$.

\end{document}